\documentclass[letterpaper,11pt]{article}
\usepackage{jheppub}
\usepackage{xcolor}
\usepackage{braket}
\usepackage{graphicx}
\usepackage{soul}
\usepackage{amssymb,amsmath}

\newcommand{\nn}{\nonumber\\}
\DeclareMathOperator{\tr}{tr}
\usepackage{verbatim}
\usepackage{anyfontsize}
\usepackage{dsfont}
\usepackage{tikz}
\usepackage{sidecap}
\usepackage{subfig}
\sidecaptionvpos{figure}{t}
\usepackage[english]{babel}
\usepackage{enumitem}  

\usepackage{graphicx, xcolor, varwidth}
\usepackage{setspace}
\usepackage[permil]{overpic}
\usepackage{graphicx}
\usepackage{mathtools}
\usepackage{wasysym}
\usepackage{graphicx}
\usepackage{placeins}
\usepackage{bm}
\newcommand{\be}{\begin{equation}}
\newcommand{\ee}{\end{equation}}

\usepackage{amsmath}
\usepackage{amsfonts}
\usepackage{amssymb}
\usepackage{graphicx}
\usepackage{wrapfig}
\usepackage{braket}
\usepackage{hyperref}
\usepackage{amsthm,verbatim,cancel}
\usepackage{color}
\usepackage{caption}
\newcommand{\bea}{\begin{eqnarray}}
\newcommand{\eea}{\end{eqnarray}}

\usepackage{tikz}
\usepackage{hyperref}
\usepackage{amssymb}
\usepackage{amsmath}
\usepackage{color}

\newcommand \mathtikz[1] {\quad \vcenter{\hbox{\tikz{#1}}} \quad}

\newcommand\Int[2]{ 
\begin{scope}[xshift=#1,yshift=#2]
\draw (-0.25,.25) -- (0.25,.25);
\end{scope}
}

\newcommand\idC[2] { 
\begin{scope}[xshift=#1,yshift=#2]
\filldraw[left color=lightgray, right color=white] (-0.25,0) -- (0.25,0) -- (0.25,-1) to [in=-90,out=-90] (-0.25,-1) -- (-0.25,0);
\filldraw[left color=white,right color=lightgray] (0,0) ellipse (0.25 and 0.1);, 
\draw[dotted] (0.25,-1) arc (0:180:0.25 and 0.1);
\end{scope}
}

\newcommand\idA[2] { 
\begin{scope}[xshift=#1,yshift=#2]
\filldraw[fill=white,draw=black] (-0.25,0) rectangle (0.25,-1);
\end{scope}
}

\newcommand\muC[2]{ 
\begin{scope}[xshift=#1,yshift=#2]
\filldraw[left color=lightgray, right color=white] (-0.25,0) to [out=-90,in=180] (0,-0.33) to [in=-90,out=0] (0.25,0) to  (0.75,0) to [in=90,out=-90] (0.25,-1) to [out=-90,in=-90] (-0.25,-1) to [in=-90,out=90] (-0.75,0);
\filldraw[left color=white,right color=lightgray] (-0.5,0) ellipse (0.25 and 0.1);
\filldraw[left color=white,right color=lightgray] (0.5,0) ellipse (0.25 and 0.1);
\draw[dotted] (0.25,-1) arc (0:180:0.25 and 0.1);
\end{scope}
}

\newcommand\pairC[2]{ 
\begin{scope}[xshift=#1,yshift=#2]
\filldraw[left color=lightgray, right color=white] (-0.25,0) to [out=-90,in=180] (0,-0.33) to [in=-90,out=0] (0.25,0) to  (0.75,0) to [out=-90,in=0] (0, -0.83) to [out=180,in=-90]
(-0.75,0);
\filldraw[left color=white,right color=lightgray] (-0.5,0) ellipse (0.25 and 0.1);
\filldraw[left color=white,right color=lightgray] (0.5,0) ellipse (0.25 and 0.1);
\end{scope}
}

\newcommand\copairC[2]{ 
\begin{scope}[xshift=#1,yshift=#2]
\filldraw[left color=lightgray, right color=white] (-0.25,0) to [out=90,in=180] (0,0.33) to [in=90,out=0] (0.25,0) to [out=-90,in=-90] (0.75,0) to [out=90,in=0] (0, 0.83) to [out=180,in=90]
(-0.75,0) to [out=-90,in=-90] (-0.25,0);
\draw[dotted] (-0.25,0) arc (0:180:0.25 and 0.1);
\draw[dotted] (0.75,0) arc (0:180:0.25 and 0.1);
\end{scope}
}

\newcommand\deltaC[2]{
\begin{scope}[xshift=#1,yshift=#2]
\filldraw[left color=lightgray, right color=white] (-0.25,-1) to [out=90,in=180] (0,-0.66) to [in=90,out=0] (0.25,-1) to [out=-90,in=-90] (0.75,-1) to [in=-90,out=90] (0.25,0) to (-0.25,0) to [in=90,out=-90] (-0.75,-1) to [out=-90,in=-90] (-0.25,-1);
\filldraw[left color=white,right color=lightgray] (0,0) ellipse (0.25 and 0.1);
\draw[dotted] (-0.25,-1) arc (0:180:0.25 and 0.1);
\draw[dotted] (0.75,-1) arc (0:180:0.25 and 0.1);
\end{scope}
}

\newcommand\muA[2]{ 
\begin{scope}[xshift=#1,yshift=#2]
\draw (-0.75,0) -- (-0.25,0) to [out=-90,in=180] (0,-0.33) to [in=-90,out=0] (0.25,0) -- (0.75,0) to [in=90,out=-90] (0.25,-1);
\draw (-0.25,-1) -- (0.25,-1);
\draw (-0.75,0) to [in=90,out=-90] (-0.25,-1);
\end{scope}
}

\newcommand\deltaA[2]{ 
\begin{scope}[xshift=#1,yshift=#2]
\draw (-0.75,-1) -- (-0.25,-1) to [out=90,in=180] (0,-0.66) to [in=90,out=0] (0.25,-1) -- (0.75,-1) to [in=-90,out=90] (0.25,0) -- (-0.25,0) to [in=90,out=-90] (-0.75,-1);
\end{scope}
}

\newcommand\zipper[2]{ 
\begin{scope}[xshift=#1,yshift=#2]
\draw (-0.25,-1) -- (0.25,-1);
\filldraw[right color=white,left color=lightgray] (-0.25,0) to (-0.25,-1) to [out=90,in=225] (0,-0.5) to [out=-45,in=90] (0.25,-1) to (0.25,0);
\filldraw[left color=white,right color=lightgray] (0,0) ellipse (0.25 and 0.1);
\end{scope}
}

\newcommand\cozipper[2]{ 
\begin{scope}[xshift=#1,yshift=#2]
\draw (-0.25,0) -- (0.25,-0);
\filldraw[right color=white,left color=lightgray] (-0.25,-1) to (-0.25,0) to [out=-90,in=135] (0,-0.5) to [out=45,in=-90] (0.25,0) to (0.25,-1) to [in=-90,out=-90] (-0.25,-1);
\draw[dotted] (0.25,-1) arc (0:180:0.25 and 0.1);
\end{scope}
}

\newcommand\epsilonC[2]{ 
\begin{scope}[xshift=#1,yshift=#2]
\filldraw[right color=white,left color=lightgray] (-0.25,0) to [out=-90,in=180] (0,-0.33) to [in=-90,out=0] (0.25,0);
\filldraw[left color=white,right color=lightgray] (0,0) ellipse (0.25 and 0.1);
\end{scope}
}

\newcommand\etaC[2] { 
\begin{scope}[xshift=#1,yshift=#2]
\filldraw[right color=white,left color=lightgray] (-0.25,0) to [out=90,in=180] (0,0.33) to [in=90,out=0] (0.25,0) to [in=-90,out=-90] (-0.25,0);
\draw[dotted] (0.25,0) arc (0:180:0.25 and 0.1);
\end{scope}
}

\newcommand\etaA[2] {
\begin{scope}[xshift=#1,yshift=#2]
\draw (-0.25,0) -- (0.25,0);
\draw (-0.25,0) to [out=90,in=180] (0,0.33) to [in=90,out=0] (0.25,0);
\end{scope}
}

\newcommand\tauA[2] { 
\begin{scope}[xshift=#1,yshift=#2]
\draw (-0.75,0) -- (-0.25,0) to [out=-90,in=90,looseness=0.5] (0.75,-1) -- (0.25,-1) to [out=90,in=-90,looseness=0.5] (-0.75,0);
\filldraw[fill=white,draw=black] (0.75,0) -- (0.25,0) to [out=-90,in=90,looseness=0.5] (-0.75,-1) -- (-0.25,-1) to [out=90,in=-90,looseness=0.5] (0.75,0);
\end{scope}
}

\newcommand\tauC[2] {
\begin{scope}[xshift=#1,yshift=#2]
\filldraw[right color=white,left color=lightgray] (-0.75,0) -- (-0.25,0) to  [out=-90,in=90,looseness=0.5] (0.75,-1) to [in=-90,out=-90] (0.25,-1) to [out=90,in=-90,looseness=0.5] (-0.75,0);
\filldraw[right color=white,left color=lightgray] (0.75,0) -- (0.25,0) to [out=-90,in=90,looseness=0.5] (-0.75,-1) to [in=-90,out=-90] (-0.25,-1) to [out=90,in=-90,looseness=0.5] (0.75,0);
\filldraw[left color=white,right color=lightgray] (-0.5,0) ellipse (0.25 and 0.1);
\filldraw[left color=white,right color=lightgray] (0.5,0) ellipse (0.25 and 0.1);
\draw[dotted] (-0.25,-1) arc (0:180:0.25 and 0.1);
\draw[dotted] (0.75,-1) arc (0:180:0.25 and 0.1);
\end{scope}
}

\begin{document}
\title{Edge modes, extended TQFT, and measurement based quantum computation}
\author{Gabriel Wong}
\affiliation{Harvard Center of Mathematical Sciences and Applications, 20 Garden Street, Cambridge MA, 02138 USA}
\affiliation{Mathematical Institute, University of Oxford
Radcliffe Observatory Quarter, Woodstock Road, Oxford OX2 6GG, United Kingdom}

\abstract{Quantum teleportation can be used to define a notion of parallel transport which characterizes the entanglement structure of a quantum state \cite{Czech:2018kvg}.   This suggests one can formulate a gauge theory of entanglement.
In \cite{Wong:2022mnv}, it was explained that measurement based quantum computation in one dimension can be understood in term of such a gauge theory (MBQC).   In this work, we give an alternative formulation of this ``entanglement gauge theory" as an extended topological field theory.   This formulation gives a alternative perspective on the relation between  the circuit model and MBQC. In addition, it provides an interpretation of MBQC in terms of the extended Hilbert space construction in gauge theories, in which the entanglement edge modes play the role of the logical qubit.    }

\maketitle

\section{Introduction}

Measurement-based quantum computation (MBQC) has emerged as a powerful framework for quantum information processing, in which quantum operations are effected through sequences of local measurements on a highly entangled resource state \cite{Raussendorf:2002zji}. 
This idea may sound counter-intuitive, since the conventional circuit model of quantum computation involves unitary evolution of a logical qubit.   However, one can connect MBQC  to the circuit model by interpreting each local measurement as implementing a gate teleportation, in which a logical qubit is simultaneously rotated and teleported.  The gate teleportation plays the role of unitary evolution in the circuit model, thus allowing measurements to induce a quantum computation.   

Beyond its computational utility, recent developments have revealed deep structural parallels between MBQC and gauge theory \cite{Wong:2022mnv}. This is because gate teleportation  defines a notion of parallel transport for the logical qubit.   Thus the teleported qubit can be viewed as a particle moving under the influence of a background gauge field.  For the case of MBQC in 1+1 D, the relation to gauge theory was clarified in \cite{Wong:2022mnv} by formulating MBQC on a circular spatial lattice.  MBQC then becomes a protocol for teleportation  ``around a closed loop", in which the logical qubit self- annhilates after being parallel transported back to its initial position. The computational output of MBQC is identified with the holonomy of the gauge field, which measures the relative rotation of the qubit due to its parallel transport.   Following \cite{Czech:2018kvg}, we refer to this transformation as an \emph{entanglement holonomy}.  This gives a gauge theoretical characterization of the entanglement structure of the resource state.

While the previous work gave an operational definition for a gauge theory underlying MBQC, this work aims to give a more formal, top-down definition using the frame work of extended topological field theory.   To motivate the introduction of this formalism, it is useful to first recall an aspect of gauge theory that is essential to our formulation of MBQC on a spatial circle.
This is the extended Hilbert space construction \cite{Donnelly:2014gva,Donnelly:2014fua,Donnelly2011}, which provides a definition of Hilbert space factorization in the presence of gauge constraints, corresponding to stabilizer constraints in MBQC. The essential idea is that in a neighborhood of an entangling surface separating spatial subregions (which are intervals in 1+1D),  the gauge constraints that cross the entangling surface must be lifted.  This leads to edge degrees of freedom at the entangling surface, which we refer to as entanglement edge modes.  The presence of these edge modes extends the Hilbert space of the gauge theory, and allows for the factorization of the gauge invariant, physical Hilbert space via the embedding:
\begin{align}\label{fact}
i:\mathcal{H}_{\text{physical}} \longrightarrow  \mathcal{H}_{I}\otimes \mathcal{H}_{I}  \qquad \qquad \vcenter{\hbox{\includegraphics[scale=.18]{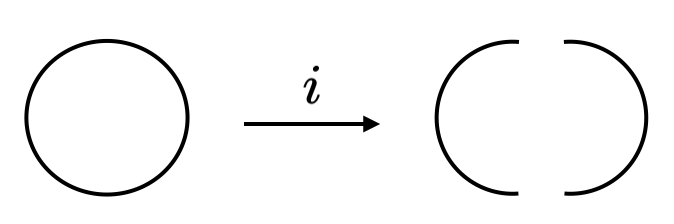}}}
\end{align} 
 The states in $\mathcal{H}_{\text{physical}} $ are labelled by fluxes, given by the holonomy of the gauge field around the spatial circle, while $\mathcal{H}_{I}$ is the gauge theory Hilbert space on an interval. 

 In the context of MBQC, $\mathcal{H}_{\text{physical}} $ plays the role of a code space, and the embedding \eqref{fact} is the encoding map satisfying the isometry condition  
 \begin{align}
     i^{\dagger} i = 1.
 \end{align} This ensures that inner products are preserved by the encoding.  By  iterating this mapping, we can embedd the code space into a lattice Hilbert space:
 \begin{align}
     \includegraphics[scale=.2]{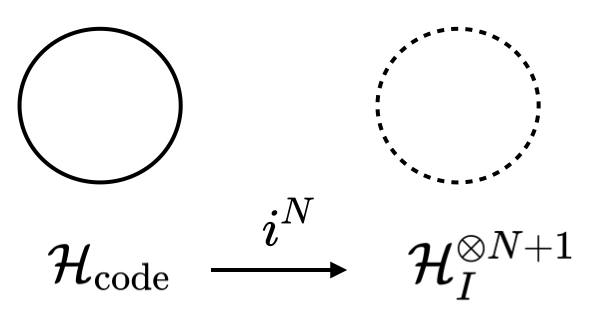}
 \end{align}
 
 The complicated features of MBQC protocol on the lattice are simplified when we consider MBQC on $\mathcal{H}_{\text{code}}$, which is spanned by states of definite ``MBQC fluxes" \cite{Wong:2022mnv}.  Moreover, the logical qubit in MBQC on which unitary gates act can be identified with the  entanglement edge modes introduced in the factorization map \eqref{fact} \footnote{ In this sense, the computational output of MBQC  provides a \emph{ non local} probe of the entanglement edge modes by measuring their entanglement holonomy.   This gives a useful operational meaning to entanglement edge modes, which  otherwise may seem like an un-physical artifice introduced for the abstract factorization of a gauge theory Hilbert space.  }.



  In \cite{Donnelly:2018ppr}, it was shown that the extended Hilbert space formalism can be most coherently formulated within the framework of extended TQFT.    In applications to MBQC, we view the TQFT as a categorical formulation of quantum mechanics that gives a useful description of quantum information processing on many body systems\footnote{This is a similar in spirit to ZX calculus \cite{Duncan:2009ocf}} .  We will show that it provides a complementary perspective to MBQC gauge theory put forth in \cite{Czech:2018kvg}.  As in the gauge theory description of \cite{Czech:2018kvg}, the TQFT language is well suited to describe global aspects of MBQC.   It also offers several additional advantages.  First, it provides  an abstract perspective that captures the relation between MBQC and the circuit model in a precise and simple manner. 
Heuristically, this relation can be captured by a single diagrammatic equation displayed in figure \ref{pch}.   Second, extended TQFT provides an \emph{algebraic} characterization of the encoding map which supports MBQC. 
Finally, the abstract TQFT language is well suited for generalization higher dimensions, where MBQC can realize universal computations.  
The following table provides a dictionary between extended TQFT and MBQC, which we will flesh out over the course of this paper. 
\\

\begin{tabular}{|c|c|}
  \hline
  \textbf{MBQC}  & \textbf{G-equivariant  Extended TQFT}  \\
  \hline
  code subspace  & Hilbert space of G-fluxes on a circle   \\\hline
  lattice Hilbert space   & extended Hilbert space   \\
  \hline
  isometric encoding map & a co-product of the Turaev algebra
  \\
  \hline
  logical qubit & entanglement edge mode \\
  \hline 
  quantum circuit/teleportation  & 1-morphism/parallel transport  \\
  \hline 
  temporally ordered measurement protocol & fusion and projection of flux eigenstates\\
  \hline computational output & holonomy around a circle\\
  \hline
\end{tabular}

\section{Review of MBQC and summary of the main ideas} \label{sec:review}
\begin{figure}[h]
    \centering
    \includegraphics[scale=.45]{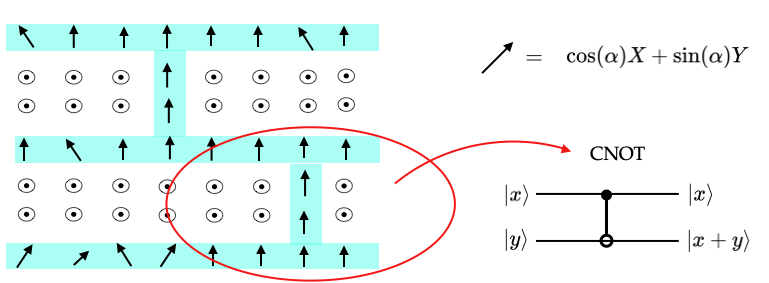}
    \caption{The left figure shows a 2D MBQC resource state-the cluster state-on which a quantum circuit has been programmed via measurements.  The symbol $\odot$ represents a measurement along $Z$, and the arrows are measurements along  $\cos \alpha X + \sin \alpha Y$. Z measurements disentangles the qubit, leaving behind a network of entangled qubits.  Measurements of this network implements gate teleportations that simulate a quantum circuit.  For example, we have circled the junction which implements the CNOT gate, defined on the right (here $x,y =0,1$ labels the states of the qubits ) }
    \label{fig:UMBQC}
\end{figure}

We begin by reviewing the basic aspects of MBQC and summarizing the main results of the paper.   The MBQC algorithm is illustrated in  figure \ref{fig:UMBQC}, which shows a 2D resource state-the cluster state-on which a quantum circuit has been programmed via measurements.  The symbol $\odot$ represents a measurement along $Z$, and the arrows are measurements along  $\cos \alpha X + \sin \alpha Y$. Z measurements disentangles the qubit, leaving behind a network of entangled qubits.  Local measurements of this network implements gate teleportations that simulate a quantum circuit, as shown in the right of figure \ref{fig:UMBQC}.
One crucial aspect of MBQC is that it involves a feed-forward mechanism: measurement angles at each time step (each column of qubits) must depend on the measurement outcomes of the previous time steps.  This temporal order determines an arrow of time on the lattice. 

Why does MBQC work? What is the origin of temporal order? We will address these questions within the framework of extended topological quantum field theory.

\subsection{MBQC in 1D}
\begin{figure}[h]
      \centering
                    \includegraphics[scale=.2]{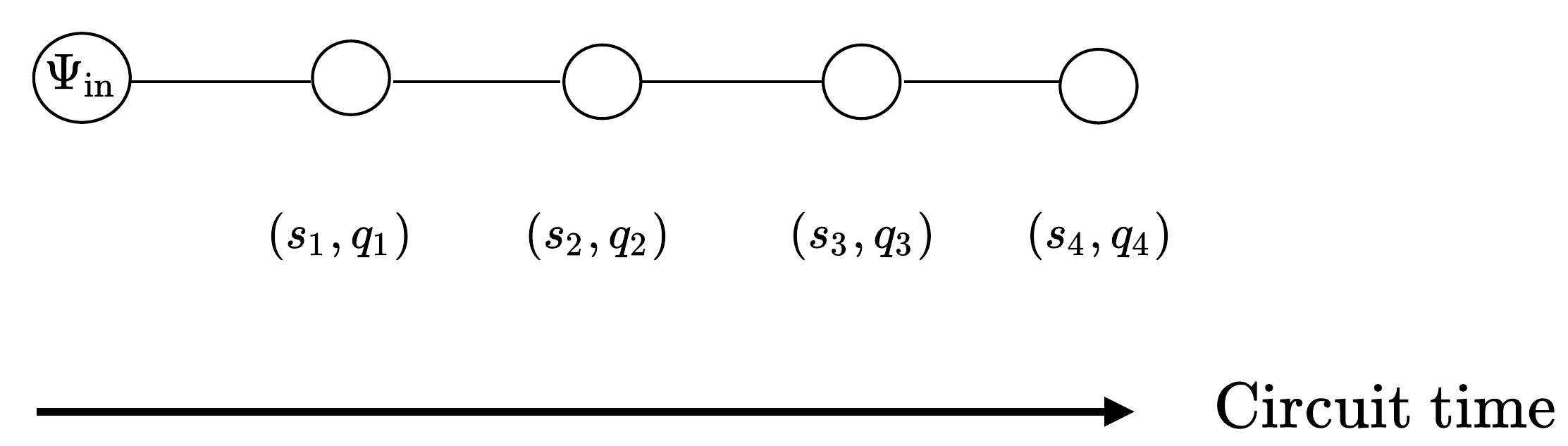}\qquad \qquad    \includegraphics[scale=.2]{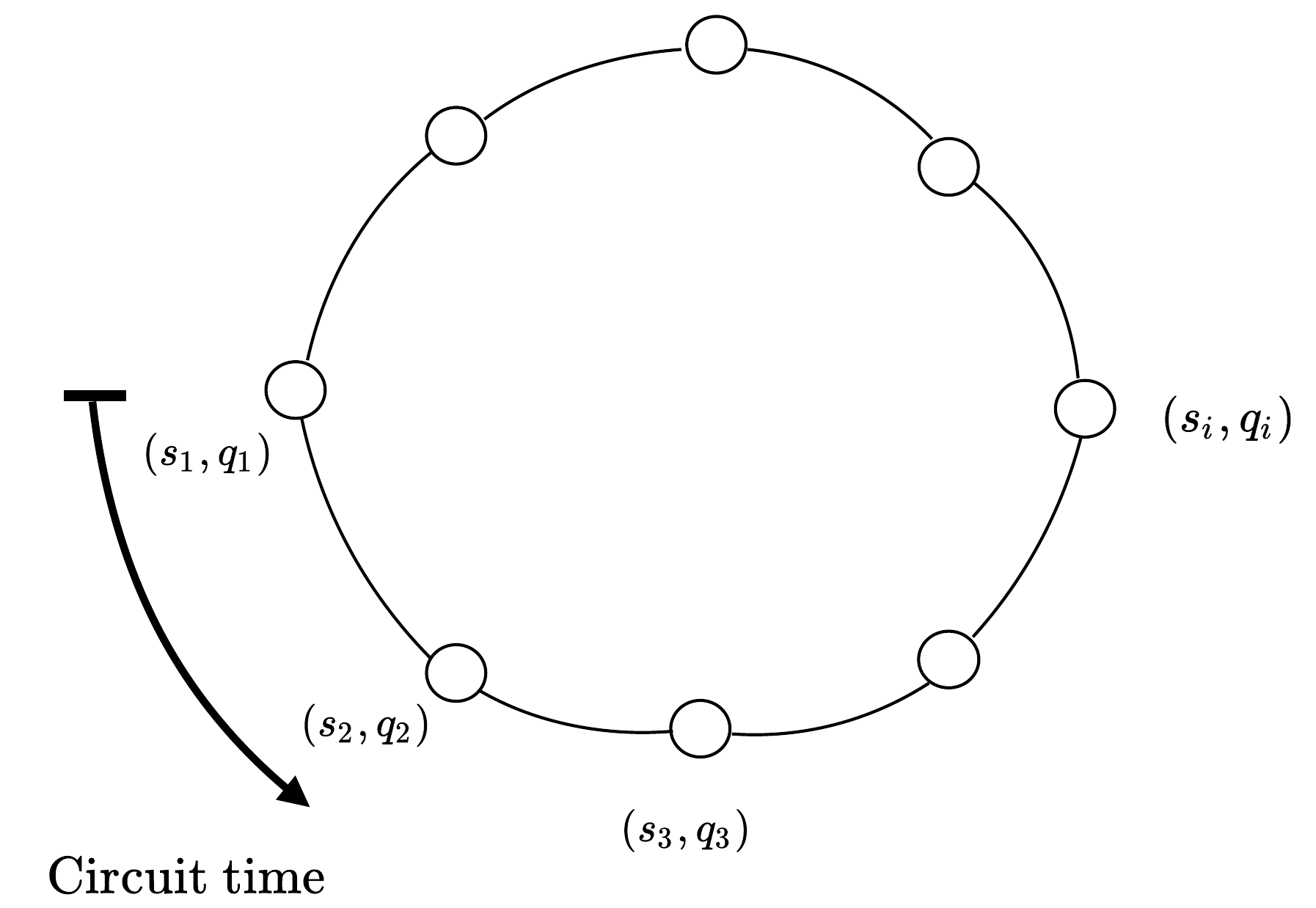}
                    \caption{The left figure illustrates the conventional 1D MBQC protocal on a chain. $\Psi_{\text{in}}$ is a logical qubit that is not measured.  After measuring the rest of the chain, $\Psi_{\text{in}}$ is teleported to the right edge of the chain, and subjected to a unitary rotation that determines the quantum computation.  On the right, we show the MBQC protocol on a circle, where all qubits are measured in a temporally ordered way. These measurements are equivalent to performing a unitary gate. However, in this case, the logical qubit is hidden.    }
                    \label{1DMBQC}
    \end{figure}

Here we review the details of the measurement protocol in the simplified setting of 1D MBQC, which is capable of simulating single qubit unitary gates.  This is illustrated in figure \ref{1DMBQC}.
The left figure shows the protocol on a 1D cluster chain with input, which is a entangled state of 2N qubits. Here, the first qubit $\Psi_{\text{in}}$ is a fixed degree of freedom which plays the role of an input that is not measured \footnote{The input qubit should be distinguished from the computational input of MBQC, which is a \emph{classical variable} }.  

 To simulate a unitary $U$, we measure the rest of the chain sequentially according to the following algorithm.

Consider a decomposition $U \in SU(2)$ into  an alternating sequence of $2N$ rotations around the $X$ and $Z$ axis:
\begin{align}\label{Ualpha}
   U(\vec{\alpha}) =\exp\left(-i\frac{\alpha_{2N}}{2}Z\right)
\exp\left(-i\frac{\alpha_{2N-1}}{2}X\right) \exp\left(-i\frac{\alpha_{2N-2}}{2}Z\right) \cdots \exp\left(-i\frac{\alpha_{2}}{2}Z\right)\exp\left(-i\frac{\alpha_{1}}{2}X\right)
\end{align}
This decomposition exists for any unitary if $N\geq 2$, since any unitary matrix U has an Euler decomposition 
\begin{align}
U=\exp\left(-i\frac{\alpha_3}{2}Z\right)
\exp\left(-i\frac{\alpha_2}{2}X\right) \exp\left(-i\frac{\alpha_1}{2}Z\right)
\end{align} 

Given the decomposition \eqref{Ualpha}, the MBQC protocol measures the following set of spins on the $X-Y$ plane:
\begin{equation}\label{orot}
O\big((-)^{q_{i}}\alpha_{i}\big) := 
\cos (\alpha_{i})X + \sin\big((-)^{q_{i}}\alpha_{i}\big) Y ,
\end{equation}
where 
\begin{equation}
\label{CPR1}
q_i =  \, \mathbf{s}_{i-1} + \mathbf{s}_{i-3} + \cdots   + \big( \mathbf{s}_2~{\rm or}~\mathbf{s}_1 \big). 
\end{equation}
Here $\mathbf{s}_{i}=0,1$  is the outcome of measuring the $i$th spin along the angle  $(-)^{q_i} \alpha_{i}$.   $q_{i}$ is part of the \emph{classical side processing data} of MBQC, and indicates which spin direction we should measure based on previous outcomes \footnote{ The purpose of this feedforward mechanism is to compensate for the random fluctuations of the measurement outcome that do not encode information. }.

 As explained in \cite{Raussendorf:2002zji}, these measurements simulate a unitary $U(\{\alpha_i\}) $ acting on $\Psi_{\text{in}}$ via  a sequence of half-teleportations that sends $\ket{\Psi_{in}}$ down the chain.  The information about the simulation is obtained from the computational output \begin{align}
     \mathbf{a}=\sum_{i\,\text{odd}} \mathbf{s}_{i} \in \mathbb{Z}_{2} 
 \end{align}
 with the sum taken mod 2. 
Specifically, the probability Prob($\mathbf{a}$) of measuring $\mathbf{a}$ determines the matrix elements of $U$:
\begin{align} \label{Uelement}
 |\langle \mathbf{a}|U|\Psi_{in}\rangle|^{2} = \text{Prob}(\mathbf{a})
 \end{align}
In \cite{Wong:2022mnv}, it was shown that to make the full gauge symmetry of MBQC manifest, we should modify this protocol by performing 1D MBQC on a cluster state living on periodic lattice (right of figure \ref{1DMBQC}).    On the circle we measure \text{all} qubits sequentially according to equation \eqref{orot} and \eqref{CPR1}.  With this modified protocol
the computational output is the $\mathbb{Z}_{2}\otimes \mathbb{Z}_{2}$ variable 
\begin{align}
g\equiv (\mathbf{a},\mathbf{b}) = (\sum_{i \, \text{odd}}  \mathbf{s}_{i}, \sum_{i\, \text{even}} \mathbf{s}_{i}) 
\end{align} 
The probability factor Prob(g) now measure the coefficients $c_{g}$ in the expansion of $U$ in terms of Pauli matrices $\{V_{g}, g \in \mathbb{Z}_{2} \otimes \mathbb{Z}_{2}  \}= \{1,X,Z,XZ\} $, viewed as a projective representation of $\mathbb{Z}_{2} \otimes \mathbb{Z}_{2} $ :
\begin{align}\label{U}
    U= \sum_{g} c_{g} V_{g}= c_{00}1 +c_{01}X +c_{10}Z+ c_{11}XZ
\end{align}
\subsection{The TQFT perspective}
The description of MBQC on a 1D chain makes manifest that the logical qubit in the computation is a physical edge degree of freedom and the unitary gates are effected through teleportation.  However 
the formulation of MBQC on a circle does not involve any boundaries, so it would seem that edge modes are not relevant.  Indeed the logical qubit is hidden in this formulation: all qubits in the physical Hilbert space of the lattice are subjected to projective measurements, so there is no unitary evolution to be found.    This obscures the relation between MBQC and the conventional circuit model.  

However, the circuit model becomes manifest when we apply the extended Hilbert space perspective to the entangled resource states for MBQC: in one spatial dimension,  these are states belonging to an SPT phase \cite{Stephen:2016vgq}.   We interpret these entangled states as a basis for code subspace of the lattice Hilbert space.   
We will identify this code space with the Hilbert space of a G-equivariant TQFT associated to a SPT phase \cite{Moore:2006dw,Shiozaki:2016cim}: heuristically this is a TQFT coupled to background fluxes in $G$.    In the TQFT language, the basis for the code space is generated by the ``cap" states with a definite flux $g \in G$
\begin{equation} 
\ket{V_{g}}=\mathtikz{ \etaC{0}{0}  ;\filldraw[red] (0,.3) circle (1pt) node[anchor=south] {\small $g$} }  ,\ g \in G
\end{equation}
We will explain how MBQC manipulates these entanglement fluxes to implement quantum computations.

Invoking the extended Hilbert space construction in MBQC may seem odd, since the lattice Hilbert space actually factorizes.  However, we would like to offer a fresh perspective on MBQC by starting with an abstract code space with no local structure. This is the TQFT Hilbert space on a circle, which has no local degrees of freedom.   The factorization map then becomes a choice.   Of course, we could factorize the code space according to the tensor product structure of the lattice.  But such a factorization does not respect the constraints imposed by the entanglement structure of the MBQC resource states. At the fixed point of the SPT phase, these are stabilizer constraints.  Alternatively, such a factorization does not respect the symmetry that protects the SPT phase.  We will show that the appropriate  factorization can be implemented using the framework of extended TQFT, and leads to entanglement edge modes that plays the role of the logical qubits for one dimensional MBQC \footnote{The relevant of edge modes for MBQC was discussed previously in \cite{Else_2012} in the context of matrix product states. Here, we interpret these as entanglement edge modes in the context of the extended Hilbert space construction }. 

\begin{figure}[h]
\centering
\includegraphics[scale=.25]{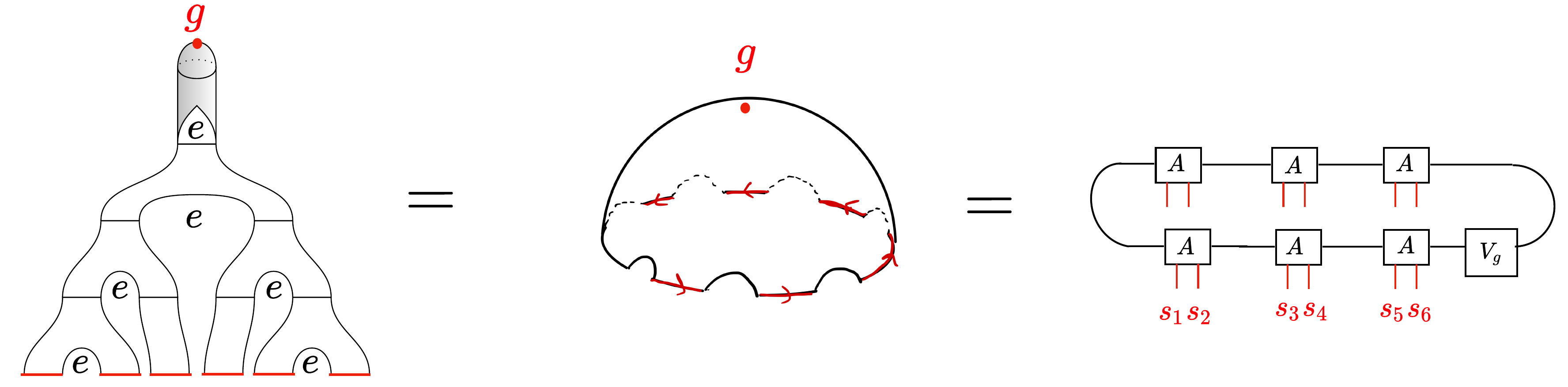}
\caption{The cap states of a G-equivariant TQFT can be encoded onto a code subspace via a Euclidean path integral process }
\label{encode}
\end{figure}

In the usual application of extended TQFT \cite{Donnelly:2018ppr}, the Hilbert space factorization is implemented by a Euclidean path integral process that cuts open a spatial slice into subregions. By iterating this factorization map, we can embedd the TQFT Hilbert space $\mathcal{H}_{S1}$ on a circle into the Hilbert space of many intervals. This gives a Euclidean spacetime interpretation to the encoding map.  This process is shown in figure \ref{encode} and forms the basis for our TQFT formulation of MBQC.  

In general, factorizing the Hilbert space via a Euclidean path integral process is meaningful because it automatically respects the constraints of locality. 
However, in the current context locality is not the issue: instead what is relevant are \emph{algebraic} constraints dictated by the entanglement structure of the resource states.   The equivalence between the local and algebraic constraints is manifest in the categorical formulation of a 2D TQFT: in fact, from this abstract perspective, a 2D TQFT \emph{is an algebra}.  Different types of TQFTs correspond to different types of algebras, each equipped with a co-product that can be used to define a factorization map.   Different factorization maps produce resource states with different patterns of entanglement.   We will show that the encoding which underlies MBQC is given by  the co-product of a Turaev algebra.  This is the symmetry algebra of the edge modes, which is preserved by the co-product.    To explain this algebraic perspective,  we will review some basic elements of category theory in sections \ref{sec:global} and \ref{sec:ETQFT}.


\subsection{Outline}
Due to the interdisciplinary nature of this work, we have included some pedagogical material, especially in extended TQFT.  In order to present a coherent story,  we have integrated our original contributions with reviews of known structures, albeit in a novel perspective.  Here we give a detailed  outline to help orient the reader. 

In \textbf{section} \ref{sec:global}, we begin with a review of closed, G-equivariant TQFT.   This provides the background needed for \textbf{section} \ref{sec:1Dsim}, where we introduce  a novel way of implementing a type of MBQC on closed TQFT states.  In particular, we emphasize the importance of the \emph{non commutative} fusion that makes MBQC possible.   

\textbf{Section} \ref{sec:ETQFT}  reviews the categorical formulation on extended TQFT and its relation to matrix product states.  This is the mathematical structure that connects the circuit model to MBQC. In particular, in \textbf{section} \ref{sec:tele} , we explain how teleportation is captured in the TQFT via morphism of the boundary category.  This mimics the way that teleportation is embedded in the MBQC protocol. 

The build up of TQFT ideas culminates in 
\textbf{section} \ref{sec:local}. where we define the isometric encoding map of figure \ref{encode} in terms  of a co-product of the extended TQFT.   This mapping is an essential element of our work, because it will be employed in section \ref{sec:PMBQC} to relate the  closed TQFT version of MBQC in section \ref{sec:1Dsim} to the actual MBQC protocol on the lattice.   In \textbf{section} \ref{sec:reference} , we review an added subtlety needed to make this connection, having to do with the choice of a reference state on the lattice \cite{Wong:2022eiu}: this is needed to define a mapping of the TQFT flux eigenstates onto the lattice Hilbert space, and will become important later when discussion measurement adaptations. 

\textbf{Section} \ref{sec:gaugesym} explains how a gauge symmetry arises from the encoding map defined in the previous section.   We show that this agrees with an operational notion of gauge symmetry as transformations that preserves the MBQC circuit simulation \cite{Wong:2022eiu}.  This makes a connection to our previous work on gauge symmetry in MBQC, and explains why the ``entanglement holonomy" that arises from teleportation around a circle agrees with the holonomy as defined in the TQFT. 

\textbf{Section} \ref{sec:PMBQC}  is the most technical section,  where we use the encoding map of figure \ref{encode} to relate fusion in closed TQFT to an operation on the lattice Hilbert space.   This fusion process is then related to the origin of the temporally ordered measurement basis in MBQC.   To make the discussion self-contained, we have included a  description of MBQC flux sectors and reference states in probabilistic MBQC. This has significant overlap with \cite{Wong:2022eiu}.

\section{A global perspective on MBQC} 
\label{sec:global}
We begin by formulating a the MBQC on a spatial circle using the language of closed TQFT.  This will highlight the global aspects of MBQC which are somewhat hidden in its usual lattice formulation. 
This protocol simulates the unitary transformation of a single logical qubit.  From the global perspective, this simulation is implemented via the measurements of ``entanglement" fluxes that thread the spatial circle \cite{Wong:2022mnv}.   As we will explain in section \ref{sec:local}, the mapping between MBQC and the circuit model requires that these fluxes take value in the projective representation of a group $G$: the theory that describes the fusion and splitting of these fluxes is given by a closed equviariant TQFT \cite{Moore:2006dw}. 

A standard interpretation of a G-equivariant TQFT is that it corresponds to coupling a TQFT  to a background flux.  Even though we will use this language to motivate the axioms of this formalism, we emphasize that in this context the background flux is really a consequence of the entanglement structure of MBQC resource states.
This perspective is manifest in the categorical formulation of TQFT that we present below, which also 
provides a useful graphical calculus to describe MBQC. To warm up to this abstract language, we begin by explaining the structure of an ordinary TQFT, prior to coupling to the background fluxes.   This will useful for highlighting the additional aspects of equivariant TQFT that makes it suitable for MBQC.   
\subsection{Ordinary closed TQFT as a commutative Frobenius Algebra}
A TQFT is an axiomatic formulation of the Euclidean path integral. In 2D, the path integral on an arbitrary closed and oriented Riemann surface can be constructed by gluing together finite set of generating surfaces:
\begin{align}\label{data}
\mathtikz{\draw[thick,->] (0,1.5) -- (0,0);
\node at (-1.5,0) {\small Euclidean time}}  \mathtikz{\muC{0}{0}},  \quad \mathtikz{\deltaC{0}{0}}, \quad \mathtikz{\etaC{0}{0}}, \quad \mathtikz{\epsilonC{0}{0}}.
\end{align}
Each of these surfaces can be regarded as a cobordism, which is  a manifold that describes the evolution between initial (top) and finial (bottom) boundaries. 
A closed TQFT translates  these path integral processes into quantum mechanical operations on a circle that define the basic data of the theory.  More precisely, a closed TQFT is rule  that
assigns to each  boundary  circle a Hilbert space $\mathcal{H}_{S^1}$, with disjoint union of circles assigned to tensor products of  $\mathcal{H}_{S^1}$ and the empty boundary assigned to $\mathbb{C}$.   Moreover, each cobordism is assigned to linear maps between the initial and final Hilbert spaces, so that the gluing of manifolds corresponds to composition of the linear maps.    

For example, the cap is a cobordism from the empty set into an oriented circle, representing a linear map
\begin{align}\label{cap}
\vcenter{\hbox{\includegraphics[scale=.6]{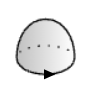}}}: \mathbb{C}\to \mathcal{H}_{S^{1}}
\end{align}
This is the preparation of a state in $\ket{\psi} \in \mathcal{H}_{S^{1}}$.    The pair of pants diagram is a linear map
\begin{align}
\mathtikz{\muC{0}{0}}: \mathcal{H}_{S^{1}}\otimes \mathcal{H}_{S^{1}}\to \mathcal{H}_{S^{1}}
\end{align} 
that defines a multiplication rule on $\mathcal{H}_{S^{1}}$.    Given a basis $\{\ket{i}, i=1 \cdots n\} $, it can be expressed in terms of fusion coefficients $c_{ij}^k$:
\begin{align}
  \ket{i}  \otimes \ket{j} \to  \sum_{k} c_{ij}^{k}\ket{k}
\end{align}  This multiplication rule makes $\mathcal{H}_{S^{1}}$ an algebra: its properties can be deduced diagramatically by imposing diffeomorphism invariance as a equivalence relation, combined with the fact that gluing of cobordisms correspond to composition of linear maps.  For example, the fact that the cap defines the unit element of the algebra is expressed by the diagrmmatic relation:
\begin{align}
\mathtikz{\etaC{-.5cm}{0} \muC{0}{0}}= \mathtikz{\idC{0}{0}},
\end{align} 
where the cylinder corresponds to the identity map.   A crucial property of the algebra is the Frobenius relation: 
\begin{equation}
\mathtikz{ \deltaC{0}{-0.5cm} \muC{0}{0.5cm} }
= \mathtikz{ \idC{4.5cm}{-0.5cm} \muC{3cm}{-0.5cm} \deltaC{4cm}{0.5cm} \idC{2.5cm}{0.5cm} }
= \mathtikz{ \idC{6.5cm}{-0.5cm}\muC{8cm}{-0.5cm}\deltaC{7cm}{0.5cm}\idC{8.5cm}{0.5cm} }.
\end{equation}

In the TQFT language,  the local properties of the path integral are captured by the fact that $\mathcal{H}_{S^1}$ is  a \emph{commutative Frobenius algebra}.  This is an algebra equipped with an invertible bilinear form 
$\eta_{ij}$ called the Frobenius form.  In the TQFT this is obtained from gluing two of the generators in \eqref{data}  
\begin{align}\label{frob}
     \vcenter{\hbox{\includegraphics[scale=.6]{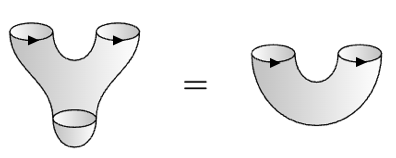}}}:  \mathcal{H}_{S^{1}}&\otimes \mathcal{H}_{S^{1}}\to \mathbb{C}\nn
                                   \ket{i} &\otimes \ket{j} \to  \eta_{ij} 
\end{align}
When the vector space $\mathcal{H}_{S^1}$ has a Hermitian inner product, the Frobenius form provides the notion of an adjoint operation that  interchanges initial and final boundaries. For example, applying it to the cap state gives its conjugate : 
\begin{align}\label{adjoint} 
\left(\vcenter{\hbox{\includegraphics[scale=.5]{OrientedCap.png}}}\right)^{\dagger} \equiv \vcenter{\hbox{\includegraphics[scale=.6]{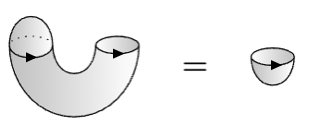}}}
  \in \overline{\mathcal{H}_{S^1}}
\end{align} 
This produces an element of the conjugate Hilbert space \footnote{In general, a Frobenius algebra is only  equipped with a vector space, with no Hermitian structure.  In this more general case, orientation reversal exchanges a state with its with its dual.} in which the action of scalar multiplication involves complex conjugation.    

A simple example of a closed TQFT is given by a $\mathbb{Z}_{n}$ valued ``matter field"  $\phi = \frac{ 2 \pi q}{n}, q=0,\cdots n-1$.  We can think of this as a theory of ground states for a scalar field subject to a strong potential $ g \cos(n x) , \quad g>>1$.  The associated Hilbert space $\mathcal{H}_{S^1}$ on a circle is just spanned by coherent states:
\begin{align}
    \{ \ket{q},q=0,\cdots{n-1}\} ,\qquad e^{i \phi } \ket{q} = e^{\frac{ 2 \pi i q}{n}} \ket{q}
\end{align}
The generators of this TQFT is given by
\begin{align}
\mathtikz{\etaC{0}{0} } &= \ket{0}\nn
\mathtikz{\epsilonC{0}{0} }&= \bra{0} \nn
    \mathtikz{ \muC{0cm}{0cm}  } &:\ket{q}\otimes  \ket{r}\to \ket{q+r} \nn
 \mathtikz{\deltaC{0cm}{0cm}}&: \ket{q} \to \sum_{r} \ket{q+r}\ket{r} 
\end{align}
The associated pairing operations
\begin{align}\label{TQFTpair}
    \mathtikz{ \pairC{0cm}{0cm}}  :\ket{q}\ket{r} \to \delta_{qr}\qquad  
    \mathtikz{ \copairC{0cm}{0cm} }  = \sum_{r} \ket{r}\ket{r}
\end{align}
define the anti linear adjoint operation:
\begin{align}
\mathcal{H}_{S^1} &\to  \overline{\mathcal{H}_{S^1}}\nn
    \ket{r} &\to \bra{r}
\end{align}

\subsection{Closed equivariant TQFT as a Turaev algebra}
We now upgrade our closed TQFT into a G-equivariant closed TQFT, which corresponds to coupling the topological matter to background fluxes labelled by elements of a group G.   In the conventional description of the path integral in terms of local fields, the standard procedure is to integrate out the matter fields, leaving behind an effective action which depends on the background field.   In the abstract TQFT formulation, we simply introduce a flat  G-gauge bundle on each boundary circle that extends into the cobordisms in \eqref{data}.   Putting a gauge bundle  on a circle introduces a holonomy $g \in G$ ,and a choice of a distinguished point $x_{0} \in S^1$.  
$$ \mathtikz{  (0,0); \draw(0,0) ellipse(.4 and 0.2) ; \filldraw[black] (0,-.2) circle (1pt) node[anchor=north]{$x_{0}$}}  $$ 

In the language of gauge fields, this is the base point which defines the path ordered exponential in the holonomy 
\begin{align}
g=P \exp \oint_{x_{0} }A                            
\end{align} 
that measures the flux threading the circle.  Since changing base points correspond to  conjugating fluxes, the base point dependence is usually removed by labelling fluxes with conjugacy classes.  However, we will see that fixing a base point is the right thing to do for describing the  MBQC protocol on a circle, since this corresponds to choosing the location of the first qubit measurement on the circle. 

The equivariant TQFT assigns to the marked circle a Hilbert space $\mathcal{H}_{S^1}$ that  decomposes into twisted sectors labelled by the flux of the gauge bundle.  This decomposition implies that $\mathcal{H}_{S^1}$ is a G-graded algebra:
\begin{align}\label{HS}
\mathcal{H}_{S^1} = \oplus_{g\in G} \mathcal{H}_{g} 
\end{align} 
Equivalently, we can view each $ \mathcal{H}_{g}$ as superselection sector corresponding to $g$ twisted boundary conditions for the matter fields.   In an \emph{invertible} equivariant TQFT, which are the TQFTs that describe SPT phases, each sector only contains a single state.   These twisted states are prepared by a cap with a puncture specifying the flux  
\begin{align}\label{cap}
\ket{V_{g}} \equiv \mathtikz{ \etaC{0}{0}  ;\filldraw[red] (0,.3) circle (1pt) node[anchor=south] {\small $g$} }  
\end{align} 
They are the relevant states needed to implement MBQC on a circle.  The reason for the notation $V_{g}$ will become apparent below.

As in the ordinary TQFT, the equivariant TQFT is determined by a finite generating set of cobordisms like \eqref{data} that connect initial and final circles with marked base points.  However, we need to introduce more data to fully specify the flat gauge bundle by indicating the holonomy along curves connecting the initial and final based points.  For example, the pair of pants cobordism is given by 
\begin{align}\label{pants} 
\vcenter{\hbox{\includegraphics[scale=.3]{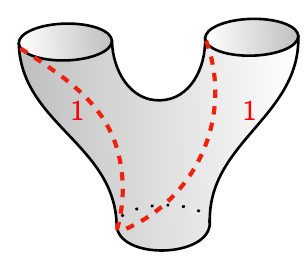}}} :\mathcal{H}_{S^1} \otimes \mathcal{H}_{S^1} \to \mathcal{H}_{S^1} 
\end{align} 
where the holonomy along the bulk curves (red dotted lines)  is set to the identity.  

In addition to \eqref{pants},\eqref{cap}, and their adjoints,   the equivariant TQFT has an additional operation given by the insertion of a nontrivial  Wilson line defect  $h$ connecting the initial and final marked points.  We can view this as a symmetry operation $\alpha_{h}$  on the twisted sectors: 
\begin{align} \label{alpha}
\mathtikz{\idC {0}{0} ;
\draw[red,->] (0,-.09) -- (0,-.7) ; \draw [red] (0,-.7) -- (0,-1.15)
; \node [red] at (.15,-.3)  {\small $h$} }  = \alpha_{h}  : \mathcal{H}_{g} \to \mathcal{H}_{hgh^{-1} } 
\end{align} 

\begin{figure}
    \centering
    \includegraphics[scale=.3]{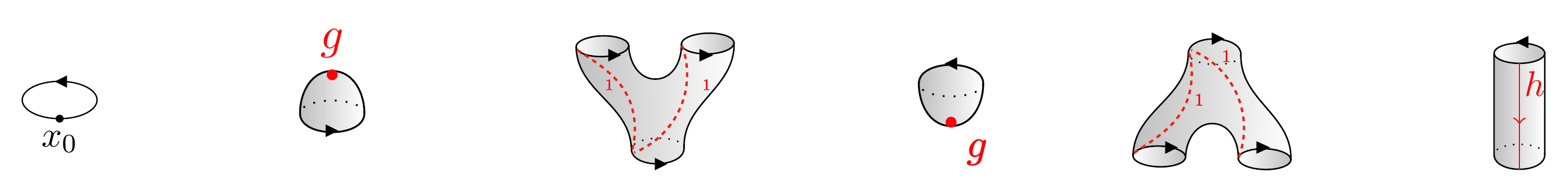}
    \caption{Cobordims of the closed G-equivariant TQFT}
    \label{fig:cobordism}
\end{figure}
As in an ordinary TQFT, the equivariant TQFT is defined by the algebraic structure on $\mathcal{H}_{S^1}$, with a multiplication defined by the fusion rule \eqref{pants}.  Crucially, this multiplication is not necessarily commutative due to the the nontrivial  composition law for the holonomies.     In fact, for invertible SPT phases,  the algebra which emerges due to the sewing relations among the cobordisms \eqref{data} and \eqref{alpha} is a special case of a Turaev  algebra \cite{Moore:2006dw}, which is classified by the second group cohomology $H^{2}(G,U(1)) $.    Thus we have the equivalence:

$$\text{Invertible Closed G-equivariant TQFT} \longleftrightarrow  H^{2}(G,U(1))    $$

We can understand this classification as follows.   $H^{2}(G,U(1))$ classifies the \emph{projective} representations of the group $G$, whose representation matrices $V_{g}$ satisfy
\begin{align}\label{rep}
V_{g_{1}}V_{g_{1}} = b(g_{1},g_{2}) V_{g_{1}g_{2}}, \qquad b(g_{1},g_{2}) \in U(1)
\end{align} 
The second cohomology classifies the possible projective phases $ b(g_{1},g_{2})$.    For each choice of  $ b(g_{1},g_{2}) \in H^{2}(G,U(1))$,  we can define a Turaev algebra via the  following product rule for fluxes:
\begin{align}\label{fusion}
\vcenter{\hbox{\includegraphics[scale=.3]{clmult.pdf}}} :  \mathcal{H}_{g_{1}} \otimes \mathcal{H}_{g_{2}} &\to \mathcal{H}_{g_{1}g_{2}} \nn
\ket{V_{g_{1}}} \otimes \ket{V_{g_{2}}} &\to b(g_{1},g_{2}) \ket{V_{g_{1}g_{2}}} .
\end{align} 
This is the algebra of the projective representation matrices.
For a general linear combination  $\alpha =\sum_{g}\alpha_{g} V_{g} $ of the representation matrices we can define an element
\begin{align}
\ket{\alpha} = \sum_{g}\alpha_{g} \ket{V_{g} }.
\end{align}
of the Turaev algebra.  For any two algebra elements $\ket{\alpha},\ket{\beta}$  equation \eqref{fusion} gives the multiplication rule
\begin{align}
    \ket{\alpha} \boxtimes \ket{\beta} = \ket{\alpha \beta } ,
\end{align}
Note that the non commutativity of this algebra is captured by the discrete torsion $\epsilon(g_{1},g_{2})$, defined by
\begin{align}
V_{g_{1}}V_{g_{2}} &= \epsilon(g_{1},g_{2})V_{g_{2}}V_{g_{1}} \nn
\epsilon(g_{1},g_{2})& \equiv \frac{b(g_{2},g_{1})}{b(g_{1},g_{2})}
\end{align} 
Thus the Turaev algebra for $G$ can be non commutative even if the group $G$ is abelian.  
As we will see below, the noncommutativity of the Turaev algebra (for nontrivial elements of $H^{2}(G,U(1))$) will be essential for the implementation of  MBQC. 

Finally, it will be useful to observe the pairing operations defined in the equivariant TQFT: 
\begin{align} \label{pairG}
    \vcenter{\hbox{\includegraphics[scale=.4]{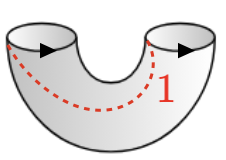}}}: \ket{V_{g}}\otimes \ket{V_{h}} \to  \delta_{gh^{-1}} ,\qquad   \vcenter{\hbox{\includegraphics[scale=.4]{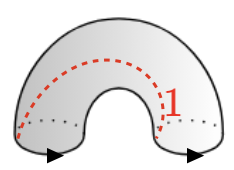}}}=\sum_{g} \ket{V_{g}} \otimes \ket{V_{g^{-1}}}
\end{align}

Note that this pairs $g$ with its inverse, which differs from the
the ordinary TQFT pairing \eqref{TQFTpair}.   This follows from the fact that when we cut open these manifolds along the dotted red lines, we obtained a simply connected space.  The total holonomy around this space must then be the identity, and this requirement determines \eqref{pairG}.  

\subsection{Simulating a single-qubit gate with fluxes. }
\label{sec:1Dsim}
The global aspects of the MBQC protocol on a circle can be beautifully captured by a  closed G-equivariant TQFT.  To be specific, let us consider $G= \mathbb{Z}_{2} \otimes \mathbb{Z}_{2} $.  This is the symmetry relevant to implementing MBQC on a 1D cluster state.  
Let us denote an element  $g \in \mathbb{Z}_{2} \otimes \mathbb{Z}_{2} $  by $g= (s_{1},s_{2}), \,\, s_{i}=0,1$, with the group law defined by addition mod 2. Then the 
nontrivial projective representation of  $\mathbb{Z}_{2} \otimes \mathbb{Z}_{2} $ is given by
\begin{align}\label{V}
V_{g}= V_{s_{1} s_{2}} = X^{s_{1}}Z^{s_{2}} .
\end{align} 
These matrices comprise the Pauli group on a single qubit.   We now give a prescription for how to implement a quantum computation in this equivariant TQFT.  
The algebra \eqref{rep} suggests that $\ket{V_{g}}$ corresponds to a quantum state in which the holonomy around the circle is $V_{g}$.    If we interpret a particle transported by these holonomies as a logical qubit, then $V_{g}$  acts as a unitary gate.    Although this idea is rather intuitive, note that we have not defined the Hilbert space of this logical qubit in the closed equivariant TQFT. 
Incorporating this logical Hilbert space into our story requires an extension of the equivariant TQFT which defines what it means to cut open the circle.   Since this requires developing more formal machinery, we defer the details to the next section.  For now, we continue with the heuristic picture that $\ket{V_{g}}$ simulates the transformation of a logical qubit by $V_{g}$.

To produce a general unitary, note that any matrix $U \in SU(2)$ can be obtained by superposing the elements of the Pauli group; 
\begin{align}\label{Uop}
U = \sum_{g} c_{g} V_{g} ,\,\,c_{g} \in \mathbb{C}
\end{align} 
If we assume linearity in the identification of states  in $\mathcal{H}_{S^1}$ with logical gates, then a general $SU(2)$ gate $U$ should be simulated by preparing the state
\begin{align}\label{U}
\ket{U}= \sum_{g} c_{g} \ket{V_{g}}=\vcenter{\hbox{\includegraphics[scale=.4]  {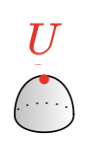}}}  
\end{align} 
This allows us to perform a simple quantum algorithm, which consists of the simulation of a unitary $U$ and then measuring it via the probability factors $|c_{g}|^{2}$. 
The computational output ( sampled from the distribution  $|c_{g}|^{2}$ ) is then given by the  $\mathbb{Z}_{2} \otimes \mathbb{Z}_{2}$ fluxes labelled by $g$. 
When $U$ is just one of the Pauli matrices $V_{g}$, this protocol is called super dense coding, which is a procedure for extracting 2 classical bits of information out of a single qubit: in our example, this is the logical qubit that is transported by holonomy $V_{g}$.   This is explained in more detail in appendix \ref{superdense}

\paragraph{Programming by measurements}
Our protocol for preparation of $U$ and its subsequent measurement can be summarized diagrammatically as follows.  We first prepare a state $\ket{U}$, which can be viewed as the result of applying an operator $U$ to $\ket{1}$, defined by left multiplication: 
\begin{align}
    \ket{U} =U \ket{1}  =\vcenter{\hbox{\includegraphics[scale=.4]{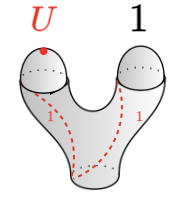} }}
\end{align}
Then we extract the computational output $c_{h}$ by  measuring the state $\ket{U}$  in the flux basis $V_{h}$. This output is described by a cobordism:
\begin{align} \label{U1h}
    c_{h} = \braket{V_{h}|U|1}= \vcenter{\hbox{\includegraphics[scale=.4]{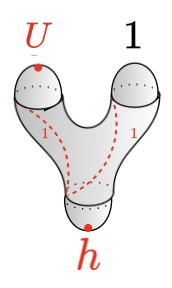}} }
\end{align}

 This differs from MBQC in a fundamental way. In our computation, the programming is implemented via the preparation of the state $\ket{U}$.  Programming thus requires  the ability to create arbitrary superpositions of the twisted states $\ket{V_{g}}$. On the other hand, MBQC is a computation protocol in which the  programming is encoded into a choice of measurement basis. The actual measurements in MBQC are local and act on single qubits on a lattice, so this measurement basis does not exist inside $\mathcal{H}_{S_{1}}$. However, as we will explain in section \ref{sec:PMBQC},  the MBQC measurements effectively projects onto basis elements in a code subspace on the lattice\footnote{A code subspace usually refers to a subspace of logical qubits.  In this context, the elements of the code subspace can be identified with coded operators via a sort of state-operator correspondence. }, and these \emph{are} represented in $\mathcal{H}_{S_{1}}$.  
 
 This TQFT basis is defined as follows.  Consider fusing both sides of equation  \eqref{U} with the flux $U^{\dagger}$.   Explicitly, if we define $U$ as 
\begin{align}
    U &= c_{1} 1 +c_{X} X +c_{Z} Z + c_{ZX} ZX
\end{align}
and 
\begin{align}
    U^{\dagger} &=  \sum_{g} \tilde{c}_{g} V_{g} =c^{*}_{1} 1 +c^{*}_{X} X +c^{*}_{Z} Z -c^{*}_{ZX} ZX,
\end{align}
then we can define 
\begin{align}\label{Udag}
     \ket{U^{\dagger}} &\equiv  \sum_{g} \tilde{c}_{g} \ket{V_{g}},
\end{align}
and a rotated flux basis: 
\begin{align}
    \ket{U^{\dagger} V_{g}} \equiv\ket{U^{\dagger}} \boxtimes \ket{ V_{g}}
\end{align}
This is the TQFT basis which corresponds to the measurement basis in MBQC.  To see why, consider fusing both sides of \eqref{U} with $\ket{U^{\dagger}}$, which  gives
\begin{align} \label{1}
   \ket{1} &= \sum_{g} c_{g} \ket{U^{\dagger} V_{g} } 
\end{align}
Here we used the identity $\ket{U^{\dagger}} \boxtimes  \ket{U} =\ket{1}$, which follows from the fact that the algebra of fluxes defined by \eqref{rep} is just the algebra of the Pauli matrices. \eqref{1} implies that we can extract the computational output $|c_{g}|^{2}$ by measuring $\ket{1}$ in the rotated flux basis $ \ket{U^{\dagger} V_{g} }$. 

From the TQFT perspective, the two measurement algorithms described above-  one using the resource state $\ket{1}$ and the other using the resource state $\ket{U}$-  are equivalent due to the adjoint identity: 
\begin{align}
    \mathtikz{\epsilonC{-.5cm}{0};\copairC{0}{0};\filldraw[red] (-.5,-.3) circle (1pt) node[anchor=north] {$U^{\dagger}$}}&= \sum_{g,h} \tilde{c}_{g}^* \braket{V_{g} | V_{h}} \otimes \ket{V_{h^{-1}}} \nn &= \sum_{h}\tilde{c}_{h}^* \ket{V_{h^{-1}}}\nn
    &=\mathtikz{\etaC{0}{0};\filldraw[red] (0,.3) circle (1pt) node[anchor=south] {$U$}},
\end{align}
where in the last equality, we used the fact that $\ket{(XZ)^{-1}} = -\ket{XZ}.$   Applying this identity to \eqref{U1h} gives two equivalent ways to measure $c_{h}$:
\begin{align}\label{pch} 
c_{h}=\vcenter{\hbox{\includegraphics[scale=.4]{U1h.png}} }=
\vcenter{\hbox{\includegraphics[scale=.25]{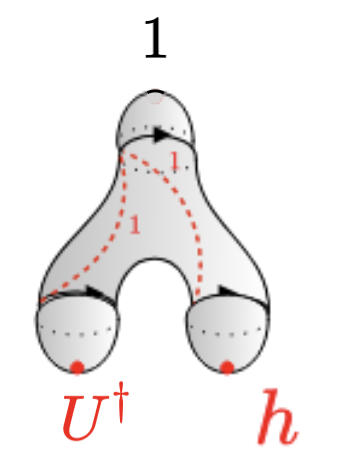}}} \longleftrightarrow \braket{V_{h}|U|1}=\braket{U^{\dagger} V_{h} | 1 } 
\end{align} 

This equivalence is a simple consequence of the fact that the equivariant TQFT faithfully captures the Hermitian structure on the algebra of Pauli matrices.  

To summarize, we have learned that there are two equivalent ways to perform a quantum computation via measurements.   In the ``active" version,  the programming is implemented by the preparation of a state $\ket{U}$  corresponding to an arbitrary $SU(2)$ rotation.    In the ``passive" version of the computation, the lab technician always prepares the state $\ket{1}$ of trivial flux \footnote{ we will later identify this state with the cluster state}, and the programming is implemented by the rotation of the measurement basis  
\begin{align}\label{rotate}
\bra{V_{g} } \to \bra{U^{\dagger}V_{g}} 
\end{align}

While these two ways of implementing the computation differ from an operational point of view, both methods work because the computational output  $|c_{g}|^{2}$ depend only on the relative rotation between the fluxes basis $\{ \ket{V_{g}}\} $ and $\{\ket{U^{\dagger} V_{g}}\}$. Naively, it may seem that neither is more advantageous than the other, since measuring in the new basis $\bra{U^{\dagger}V_{g}} $  maybe equally complicated as superposing $\ket{V_{g}}$ to create the resource state $\ket{U}$. 
However, MBQC provides an algorithm  that effectively implements measurements of the code subspace in the basis $\bra{U^{\dagger}V_{g}}$, without the need to perform arbitrary superpositions.  This drastically reduces the operational costs of programming and makes MBQC a useful protocol. 

To give an explicit description of MBQC, we will first map the TQFT states $\mathcal{H}_{S^1}$ into entangled states on a lattice.
In section \ref{sec:PMBQC}, we will show that measurements in the basis $\bra{U^{\dagger}V_{g}}$ can then be implemented by a temporally ordered sequence of single qubit measurements in which the choice of measurement angles at each step depends on the measurement outcomes of the previous steps.    Our main goal in the following sections is to explain this MBQC measurement protocol from the point of view of extended TQFT. 
\paragraph{Commutative vs Non-commutative algebras} Finally, before moving on, we emphasize that even though the computational output of MBQC is a classical variable valued in $\mathbb{Z}_{2} \otimes \mathbb{Z}_{2}$,  it is important for the purposes of the circuit simulation that the ``MBQC  fluxes"  are valued in a nontrivial \emph{projective} representation, as indicated by the non-commutative algebra \eqref{rep}.  An ordinary representation of an abelian group like $\mathbb{Z}_{2} \otimes \mathbb{Z}_{2}$  is one dimensional, and the corresponding holonomies cannot produce a nontrivial gate.   Such representations would appear in ordinary $\mathbb{Z}_{2} \otimes \mathbb{Z}_{2}$ Dijkgraaf Witten theory,  which is a  \emph{commutative} algebra as supposed to the non commutative algebra of the equivariant theory.

Notice that in the theory with a commutative algebra of fluxes, we can still form the superposition of flux eigenstates $\ket{g}_{\text{abelian}}$ given by 
\begin{align}\label{abelian}
    \ket{U}_{\text{abelian}}= \sum_{g\in G} c_{g} \ket{g}_{\text{abelian}}
\end{align}
However, the adjoint operation of the ordinary TQFT does not capture the Hermitan structure of the Pauli matrices.  For this reason, the analogue of \eqref{pch} , which is the essential ingredient for the programming of $U$ in the MBQC protocol, fails to be satisfied.   Thus, the labelling of the state in \eqref{abelian} by the operator $U$ is misleading, since the commutative algebra of the states $\{\ket{g}\}$ fails to simulate the algebra of nontrivial gates.

An alternative viewpoint, which we will illustrate in the section \ref{sec:ETQFT}, is that MBQC fails to work in ordinary TQFT's with abelian algebras because they don't have the necessary entanglement edge modes.  
\section{Edge modes and extended TQFT}
\label{sec:ETQFT}
As alluded to earlier, to give an explicit description of the circuit simulation, we need to provide a TQFT definition of the Hilbert space for the logical qubit, which is distinct from $\mathcal{H}_{S^1}$.    Intuitively we want to cut open the circle into an interval, and construct a suitable \emph{edge mode} Hilbert space at the endpoints where the logical qubit lives.   Moreover, where as cobordisms of the closed TQFT carries a notion of Euclidean time that flows from the initial to final boundary (from top to bottom),  the circuit model we are simulating requires a notion of circuit time that evolves the qubit along a spatial slice from left to right.  All of these ideas can be  incorporated into a coherent framework called \emph{extended} TQFT.   In particular, the sewing rules of the extended equivariant TQFT will link the algebra \eqref{rep} of fluxes to the algebra of representation matrices $V_{g}$ acting on the edge modes.    This defines a sort of state-operator correspondence that explains how a circuit simulation can be encoded into the twisted states \eqref{cap} of a closed equivariant TQFT.

\subsection{Category description of an open-closed TQFT}
To appreciate the logic behind extended TQFT, it is useful formulate our closed TQFT in the language of category theory.    A category consists of a set of objects and a set of relations between these objects called morphisms, which compose in an associative fashion.  In particular, the geometric category $\mathcal{C}$ of closed, 1 dimensional manifolds has circles as the objects and cobordisms between circles as the morphisms.  The composition of these morphisms correspond to the gluing of cobordisms.     In this language, a 2D closed TQFT is a map:
\begin{align}
Z_{\text{closed}} : \mathcal{C} \to \mathbf{Vect}_{\mathbb{C}}
\end{align} 
from $\mathcal{C}$ into the category  $\mathbf{Vect}_{\mathbb{C}}$  of complex vector spaces. The morphisms of  $\mathbf{Vect}_{\mathbb{C}}$ are given by linear maps  with the usual composition rule.    The fact that $Z_{\text{closed}}$ maps the gluing of cobordisms  to composition of linear maps means that $Z_{\text{closed}}$ is a \emph{functor} from $\mathcal{C}$  to  $\mathbf{Vect}_{\mathbb{C}}$.  The functor $Z_{\text{closed}}$  is therefore a linear representation of 2D manifolds and describes their gluing and cutting along co-dimension 1 manifolds.  For applications to physics, we will also require a Hermitian inner product, which defines an adjoint operation on the morphisms of $\mathcal{C}$.  In this case $\mathcal{C}$ is the category of Hilbert spaces, and $Z_{closed}$ is a ``dagger" category \cite{Abramsky:2004doh}. 

A 2D extended TQFT generalizes this categorical description of the path integral to allow for cutting and gluing along codimension 2 manifolds, which in this case is just a point.    It introduces two new ingredients, both of which are relevant to MBQC.    First, the extension of a  closed TQFT $Z_{\text{closed}}$ enlarges  the source category $\mathcal{C}$ by including  labelled intervals $\hspace{-0.25cm}\mathtikz{\Int{0}{0}; \node at (-.35,.22) {\small $a$};\node at (.35,.25)  {\small $b$}}\hspace{-0.25cm}$  as objects.  We can view the labels $a,b$  as abstract boundary conditions.     The extended TQFT assigns a Hilbert space $\mathcal{H}_{ab}$ to these intervals, which contains edge mode degrees of freedom determined by $a,b$.  The set of cobordisms is then enlarged to include manifolds with corners, which have intervals as well as circles as boundaries.   In particular, there is a fusion of intervals
\begin{align}\label{openfuse}
\mathtikz{\muA{0}{0};\node at (-.75,-.5) {\small $a$};\node at (.9,-0.5) {\small $c$};\node at (0,0) {\small $b$}} : \mathcal{H}_{ab}\otimes \mathcal{H}_{bc} \to \mathcal{H}_{ac}
\end{align} 
 which defines the multiplication rule for states in the interval Hilbert space $\mathcal{H}_{\text{open}} $:
\begin{align}
\mathcal{H}_{\text{open}} = \oplus_{a,b} \mathcal{H}_{ab} 
\end{align} 
We refer to $\mathcal{H}_{\text{open}} $ as the open algebra.   The open sector of the TQFT is naturally suited to describe MBQC on a chain.

For ordinary TQFT (in the absence of symmetry), the open algebra generically satisfies a different multiplication rule than closed algebra: in particular, the open algebra can be non commutative, whereas  the closed algebra is always commutative.  
\begin{align}
\mathtikz{ \muA{0cm}{0cm} } \neq 
\mathtikz { \muA{0cm}{-0.5cm}  \tauA{0cm}{0.5cm} },\qquad \mathtikz{ \muC{0cm}{0cm} } = \mathtikz { \muC{0cm}{-0.5cm}  \tauC{0cm}{0.5cm} }
\end{align} 
Even though they are generically not identical,  the closed and open algebra are related  by the  ``zipper" cobordism 
\begin{align}
\mathtikz{\zipper{0}{0} ; \draw (0cm,-.85cm) node { $a$}; } :\mathcal{H}_{S^1} \to \mathcal{H}_{aa}
\end{align} 
which cuts open the circle into an interval.    The sewing relations of the extended TQFT imply that this is an algebra homomorphism from the closed to the open sector.  Such a homomorphism must satisfy a set of open-closed sewing constraints \cite{Moore:2006dw}.   The solutions to these constraints represent different ways of extending the closed algebra into an \emph{open-closed} algebra that unifies both sectors.   In particular,  for an ordinary TQFT, the open-closed algebra is a \emph{knowledgeable Frobenius algebra}
\paragraph{Edge modes and the boundary category}
The second new ingredient introduced by the extended TQFT is the assignment of a \emph{boundary category} $\mathcal{B}$\footnote{Note that  $\mathcal{B}$ is a target category.  i.e the extended TQFT includes a map $Z: \text{point} \to \mathcal{B}$}  to codimension 2 surfaces which make up the endpoints of an interval.    We will choose
$\mathcal{B}$ to be the category of boundary conditions, whose objects are given by the abstract boundary  labels $a,b,\cdots$.   
Morphisms between boundary conditions $a$ and $b$ are  boundary condition changing operators $\hat{O}_{ab}$:  such operators always form the structure of a vector space (see  \cite{Kapustin:2010ta} for a nice explanation of this fact).  Concretely, in our discussion of open TQFT, these boundary condition changing operators can be identified with the elements of the interval Hilbert space $\mathcal{H}_{ab}$, which we now view as a space of morphisms between $a$ and $b$:
\begin{align}
\mathcal{H}_{ab} = \text{Hom}(a,b)
\end{align} 
The fact that elements of  $\mathcal{H}_{ab}$ are morphisms simply means that they can be composed in an associative fashion.       Their composition law is given by the open fusion rule \eqref{openfuse}, which satisfies associativity as one of the sewing relations. 

The morphisms between objects at the endpoints of an interval represent a type of ``horizontal"  evolution.   This is the TQFT formulation of circuit time evolution.  Meanwhile, the identification of these horizontal morphisms with quantum states suggests a type of quantum teleportation in which an entangled state on an interval can be used to teleport a qubit between its endpoints.   
Below, we will make these suggestive ideas precise in the context of extended equivariant TQFT.
\subsection{G-equivariant open-closed TQFT and matrix product states}

We now consider the extension of  a G-equivariant TQFT.  This again introduces edge modes in an open sector in which intervals are the objects.   These intervals are marked with a based point at the left endpoint.       As in the closed sector, all cobordisms are equipped with a flat gauge bundle, which is specified by indicating the holonomy between marked points on the initial and final intervals
 \begin{align}\label{open}
 \vcenter{\hbox{\includegraphics[scale=.4]{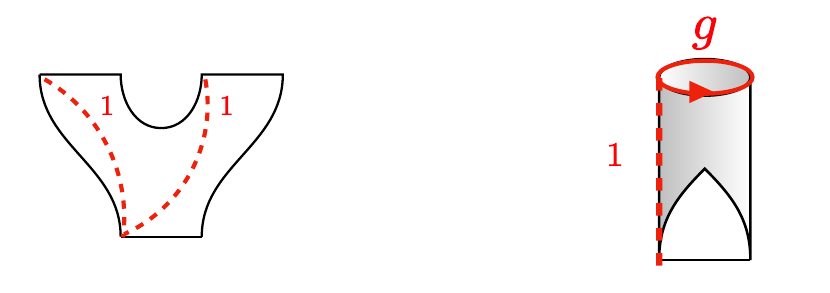}}}
  \end{align} 
In addition, the open equivariant TQFT also includes a symmetry operation: . 
 \begin{align}\label{stripsym} 
 \vcenter{\hbox{\includegraphics[scale=.1]{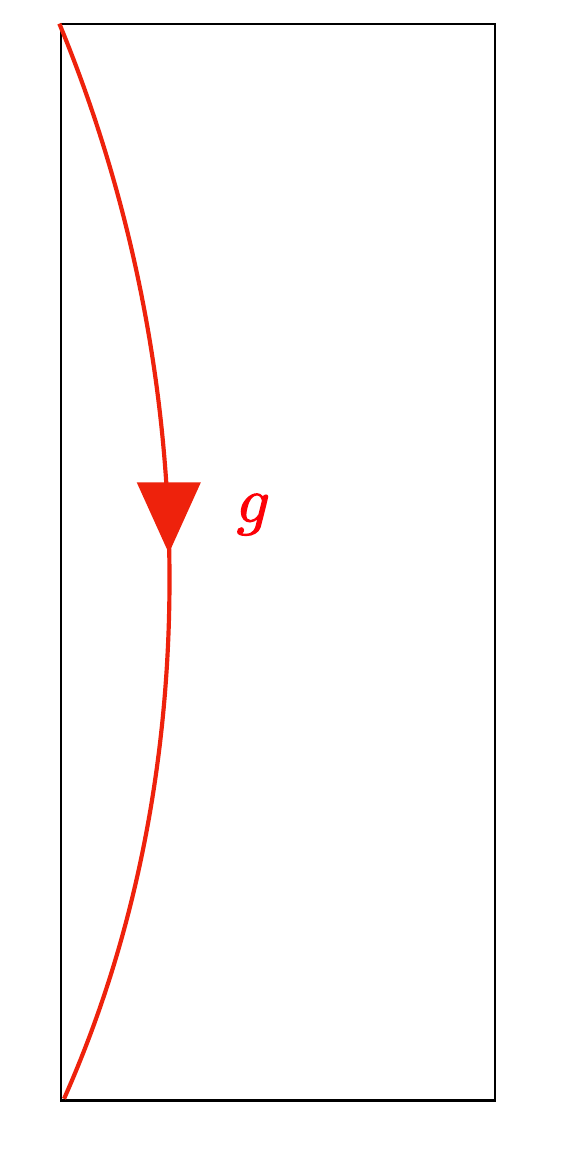}}} :\mathcal{H}_{ab} \to \mathcal{H}_{ab} 
 \end{align} 
The sewing relations for an open-closed equivariant TQFT were solved in \cite{Moore:2006dw}.     In particular, the interval Hilbert space is given by 
\begin{align} \label{Hab}
\mathcal{H}_{ab}=\mathcal{V}^{*}_{a} \otimes \mathcal{V}_{b} = \text{Hom}(a,b)
\end{align} 
where $\mathcal{V}_{a}$ are projective representations of $G$.      These projective representations are the edge modes of the G-equivariant TQFT \cite{Moore:2006dw}.   Notice that the structure of  the Hilbert space  \eqref{Hab} naturally lends itself to the interpretation as the space of morphisms between $\mathcal{V}_{a}$ and $\mathcal{V}_{b}$, since  $\mathcal{V}^{*}_{a} \otimes \mathcal{V}_{b} $ is isomorphic to the space of linear maps between $\mathcal{V}_{a}$ and $\mathcal{V}_{b}$.  This is made explicit by choosing a basis $\ket{a,i} \in \mathcal{V}_{a}$, $\ket{b,j} \in \mathcal{V}_{b}$ (we emphasize these states are distinct from states in $\mathcal{H}_{ab}$).   Then a basis for $\ket{\bar{i},j} $  for $\mathcal{H}_{ab}$ is in one to one correspondence with operators from $\mathcal{V}_{b}$ to $\mathcal{V}_{a}$:
\begin{align}\label{ij}
\ket{\bar{i},j} \equiv \ket{b,i} \otimes \ket{a,j} \leftrightarrow   \ket{b,i} \otimes \bra{a,j}   \quad i=1,\cdots \dim \mathcal{V}_{a},\quad j=1,\cdots \dim \mathcal{V}_{b},
\end{align} 
and the open string fusion corresponds to the composition of these  operators. The open algebra is thus a matrix algebra:
\begin{align}\label{omu}
\mathtikz{\muA{0}{0}} :  \ket{i,\bar{j} }  \otimes  \ket{k,\bar{l} } \to \delta_{kj}  \ket{ i \bar{l}} 
\end{align} 

\paragraph{Matrix product states}
For applications to MBQC, we will be interested in G-equivariant TQFT's that describe the RG fixed point of an SPT phase.  
In this context, the interval Hilbert space  $\mathcal{H}_{ab}$ is a coarse grained description of a 1D system with edge modes transforming in the same projective representation \footnote{  To simulate an arbitrary single qubit unitary actually requires a maximally non commutative projective representation, in which the phases $b(g,h)$ satisfy $b(g,h)  \neq \b(h,g)$  when $g\neq h$ \cite{Else_2012}} on both boundaries.  In this case, we set $a=b=\mathcal{V}$ and denote  the interval Hilbert space by
\begin{align}\label{VV}
\mathcal{H}_{\text{open}} = \mathcal{V}^* \otimes \mathcal{V} 
\end{align}  
Each quantum state in $ \mathcal{H}_{\text{open}} $ is then identified with a linear map of $\mathcal{V}$ onto itself,  which describes a quantum evolution from the left endpoint to the right endpoint of the interval.   
In particular, we will see that when $G= \mathbb{Z}_{2} \otimes \mathbb{Z}_{2}$, this horizontal evolution can be interpreted as the standard teleportation of a qubit in $\mathcal{V}$ from one endpoint to the other. 

To see the relation to teleportation, it is useful to formulate the open TQFT in terms of matrix product state (MPS) tensors.   
In this context MPS tensors arise as intertwiners\footnote{An intertwiner is a linear map between representations that commute with the action of G} between two projective representations of $G$ with the same projective phase $b(g,h)$.  One is given by $\mathcal{H}_{\text{open}} = \mathcal{V} \otimes \mathcal{V}^* $, and the other is the $b$ twisted regular representation of $G$ \cite{CHENG2015230}.
The latter can be defined abstractly via the action of $G$ on the twisted group algebra  $\mathbb{C}_{b}[G]$.  As a vector space, this algebra is spanned by a basis $\ket{g}$ labelled by group elements.  Each element in $ \mathbb{C}_{b}[G] $ is thus a formal linear combination
\begin{align}
\ket{f} = \sum_{g \in G} f(g) \ket{g}
\end{align} 
where $f(g)$ is a complex valued wavefunction on the group.
The action of an element $h\in G$ in the regular representation is defined to be
\begin{align} \label{lh}
 l(h)   \ket{g}= b(h,g) \ket{hg}
\end{align}
which corresponds to left multiplication by $\ket{h}$ in the $b$ twisted group algebra.

The projective Peter Weyl theorem \cite{cheng2015character} implies that for every nontrivial projective representations $\mathcal{V}$  of $G$, its representation matrices $V_{g}^{ij}$ define an intertwiner\footnote{This map is one to one if there is a single projective irrep. associated to the phase $b(g,h)$.  In general the Peter Weyl theorem implies that 
\begin{align}\label{PW}
  \mathbb{C}_{b}[G] =\oplus_{R} \mathcal{V}_{R}\otimes \mathcal{V}_{R}^{*} 
\end{align}
where $R$ labels the different irreps with the same phase.  In this general setting, the intertwining map between the two sides of \eqref{PW} will involve a direct sum of representation matrices.  }
\begin{align}\label{Int}
A &:\mathcal{V} \otimes \mathcal{V}^*  \to \mathbb{C}[G] \nn
A &=  \sum_{i,j} \sum_{g \in G}  V_{g}^{ij}   \ket{g}  \bra{\bar{i}\,j}
\end{align}

The interwiner $A$ can be represented graphically as a matrix product state tensor 
\begin{align}\label{MPS}
 \mathtikz{  \draw (0,0) rectangle (.7,.4);\draw (-.4,.2) -- (0,.2) ; \draw (.7,.2) -- (1.1,.2) ; \draw (.2,0)--(.2,-.3);\draw (.5,0)--(.5,-.3);\draw (0.33cm,0.20cm) node { $A$};\draw (-.6,.2) node { \footnotesize $i$};\draw (1.3,.2) node { \footnotesize $j$};\draw (.35,-.5) node { \footnotesize $g$}}
\end{align} 
We will refer to $g$ as the ``physical index" and $i,j$ as the edge mode index.   In the group basis $\ket{g}$, also called the \emph{symmetry basis} , this tensor is given by $A_{g}^{ij}= V_{g}^{ij} $.  However, to implement the MBQC protocol, we will need to project $A$ onto a more general class of measurement basis that is also labelled by group elements. For this reason, we will reserve the symbol $A_{g}^{ij}$ to denote the components of $A$ in a general basis.   In MPS notation, the intertwining property of $A$ (with $\mathcal{V}\otimes \mathcal{V}^*$ viewed as a representation of $G$ in the left tensor factor) implies the transformation law
\begin{align}\label{lg}
\includegraphics[scale=.2]{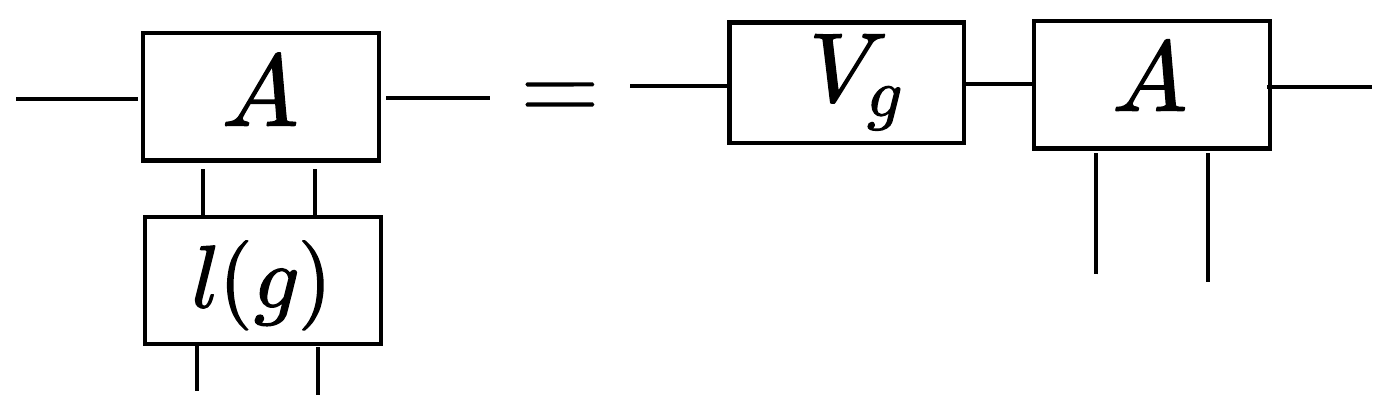} 
\end{align} 
This follows directly from the representation property \eqref{rep} of the matrices $V_{g}$.   This equation shows how the left-regular representation is equivalent to the \emph{edge modes symmetry } acting on $\mathcal{V}$.
Similarly, there is a right-regular representation $r(h)$ which acts on the right by 
\begin{align}\label{rg}
 r(h)   \ket{g}= b(g,h) \ket{gh^{-1}}
\end{align}
with the associated transformation law 
\begin{align}
\includegraphics[scale=.1]{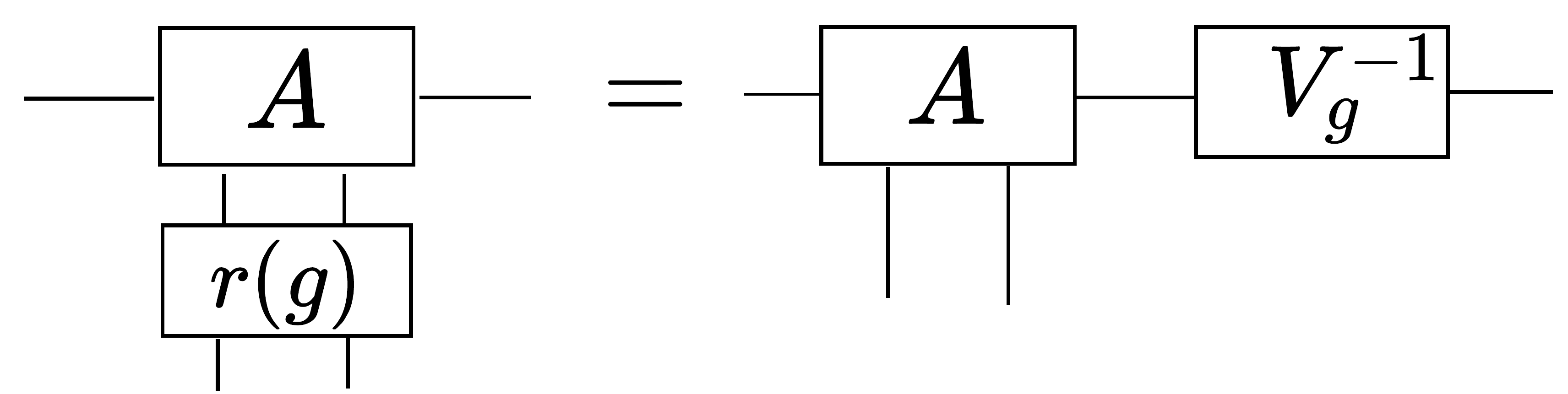} 
\end{align} 

Usually, an MPS tensor is contracted with physical basis states $\ket{g}$ to define a quantum state.  For example, for each choice of $i,j$ we obtain a state on $\mathcal{H}_{open}$ given by
\begin{align}\label{Aij}
\ket{A_{ij}} &= \sum_{g} \braket{ i| A_g|j}  \ket{g}  \in \mathcal{H}_{\text{open}} 
\end{align} 
However, to relate our discussion of extended TQFT to teleportation, it is more useful to view  $A$ as a tensor that assigns a linear map (a morphism) $A_{g}: \mathcal{V} \to \mathcal{V}$ to each quantum state $\ket{g}$.   Such a tensor can be viewed as a fluctuating quantum circuit on $\mathcal{V}$, which we can write as:
\begin{align}\label{Ag}
A= \sum_{g}  A_{g} \ket{g}  ,\qquad A_{g}: \mathcal{V} \to \mathcal{V}.
\end{align} 
In this formula, we read $A_{g}$ as an operator valued coefficient, so that a measurement that projects $A$ on to $\ket{g}$ implements the linear map $A_{g}$.  In MPS notation, we express this linear map  as:
\begin{align}\label{rcirc}
 \mathtikz{  \draw (0,0) rectangle (.7,.4);\draw (0,-.3) rectangle (.7,-.7);\draw (-.4,.2) -- (0,.2) ; \draw (.7,.2) -- (1.1,.2) ; \draw (.2,0)--(.2,-.3);\draw (.5,0)--(.5,-.3);\draw (0.33cm,0.20cm) node { $A$};\draw (.35,-.5) node { \footnotesize $g$}} : \mathcal{V} \to \mathcal{V}.
\end{align}

\subsection{Teleportation by symmetry} \label{sec:tele}
\begin{figure}[h]
\centering
\includegraphics[scale=.23]{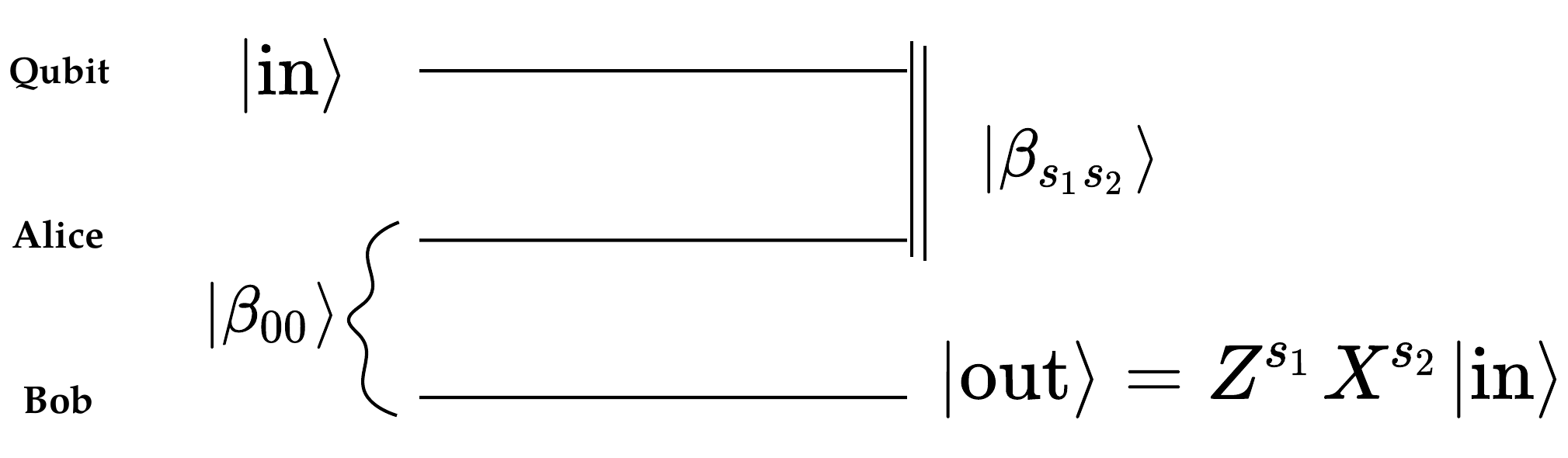}
\caption{Circuit diagram for quantum teleportation.}
\label{tel} 
\end{figure}

We now give a circuit model description of the fluctuating operator  $A$ which shows that it describes the teleportation of edge mode qubits in $\mathcal{V}$\cite{Wong:2022mnv}.   The conventional circuit diagram for teleportation is shown in figure \ref{tel}.

To teleport a qubit $\ket{\text{in}}$ from Alice to Bob, they must share an entangled resource state given by the bell pair $\beta_{00} \equiv \frac{1}{\sqrt{2}} (\ket{11}+\ket{00})$.      Alice then makes a joint measurement on herself and $\ket{in}$ qubit in a bell basis 
\begin{align}\label{bell}
\ket{\beta_{s_{1}s_{2}} }= X^{s_{2}}Z^{s_{1}} \ket{\beta_{00}} 
\end{align} 
which teleports the qubit to Bob.   Due to the randomness of Alice's measurement outcomes,  The actual qubit received by Bob is rotated by a random \emph{by product operator} $ Z^{s_{1}}X^{s{2}}$.  Thus, the teleportation protocol is captured by a random circuit of the form \eqref{Ag}, in which a measurement outcome $(s_{1},s_{2})$ implements the unitary 
\begin{align}
    A_{s_{1}s_{2}} = \mathtikz{  \draw (0,0) rectangle (.7,.4);\draw (0,-.3) rectangle (.7,-.7);\draw (-.4,.2) -- (0,.2) ; \draw (.7,.2) -- (1.1,.2) ; \draw (.2,0)--(.2,-.3);\draw (.5,0)--(.5,-.3);\draw (0.33cm,0.20cm) node { $A$};\draw (.35,-.5) node { \footnotesize $s_{1}s_{2}$}}  = Z^{s_{1}}X^{s{2}}
\end{align}
If we identify measurement outcomes $(s_{1},s_{2})$ in the bell basis with group elements:
\begin{align}
    g=(s_{1},s_{2})
\end{align}
then the teleportation protocol can be identified with the fluctuating circuit \eqref{Ag} defined by the intertwiner $A_{g}$.  More precisely, this mapping works because equation \eqref{bell} shows that the bell basis forms an orbit under the projective representation \eqref{V} of $G= \mathbb{Z}_{2} \otimes \mathbb{Z}_{2}$.  This defines a group label $g= (s_{1},s_{2}) $ for elements of this orbit with respect to the reference state $\ket{\beta_{00}}$, which allows an identification 
\begin{align}
    \ket{g} \to V(g) \ket{\beta_{00}} 
\end{align}
between the an abstract state $\ket{g} \in \mathbb{C}_{b}[G]$ and the bell basis.   

\begin{figure}
\centering
\includegraphics[scale=.23]{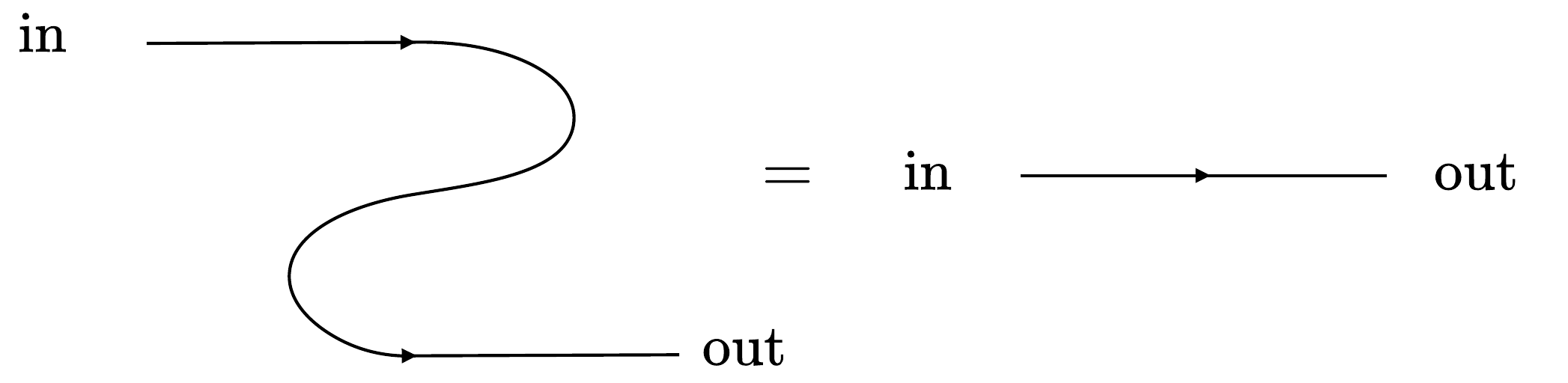}
\caption{The bell pair creation and bell measurement in the teleportation protocol corresponds to the pairing and co-pairing cobordisms of a 1D TQFT.    The zigzag identity expresses the fact that these operations combine to form the identity map.  }
\label{1Dtel} 
\end{figure}

The symmetry that relates the measurement outcomes also allows Bob to mitigate the randomness of the circuit:  if Alice communicates to her measurement outcomes to  Bob  via classical channels, he can apply an inverse operation that undoes the effects of the byproduct, thereby retrieving the original qubit.  In this sense, teleportation can be also be viewed as implementing the identity gate.  This perspective is captured by the  ``zigzag" identity\footnote{The appearance of this 1 D TQFT is not a coincidence.  There is another formulation of 2D extended TQFT in which  the interval itself is viewed as a ``1-morphism", i.e. a cobordism between the two endpoints.  In this language cobordisms between  intervals are 2-morphisms ( a morphism between morphism).  In a sense the extended TQFT provides rules that tie together TQFT in different dimensions.  In particular the zigzag identity is a sewing relation of the 1D TQFT that is refered to as ``Dualizability" } of a 1D TQFT, as illustrated in figure \ref{1Dtel}.  

\paragraph{The closed to open map}
In section \ref{sec:1Dsim}, we had claimed that  holonomy $g$ that labels the states  $\ket{V_{g}}  \in \mathcal{H}_{S^1}$ of the closed TQFT can be identified with a quantum circuit $V_{g}$ that transports a virtual particle around the circle.   Our discussion above shows that this quantum circuit is a generalized form of teleportation acting on edge modes that transform under  a projective representation of $G$.

In the language of extended TQFT, the identification between the twisted state $\ket{V_{g}}$ and the teleportation of edge modes is given by the zipper cobordism, which maps the states on the circle to those on the interval:
\begin{align} \label{clopen} 
\mathtikz{\zipper{0}{0}} : &\mathcal{H}_{S^1} \to \mathcal{H}_{\text{open}} \nn
&\ket{V_{g}} \to \ket{g}\equiv \sum_{ij} \sqrt{\dim V}\, V_{g}^{ij}  \ket{\bar{i} j}
\end{align} 
Applying the state-channel duality \eqref{ij} to the basis states $ \ket{\bar{i} j}$ then gives the identification of $\ket{V_{g}}$ with the quantum circuit $V_{g} : \mathcal{V} \to \mathcal{V} $, which is the same linear map defined by the MPS tensor in \eqref{rcirc}.   This is the state-operator correspondence we alluded to at the beginning of section \ref{sec:ETQFT}.

We can check explicitly that \eqref{clopen} is an algebra homomorphism: using the fact that the open algebra is the matrix algebra of the Pauli matrices, we apply the representation property of these matrices \footnote{Explcitly we have
\begin{align}
\mathtikz{\muA{0}{0}}: \ket{g_{1}} \otimes \ket{g_{2}} &\to  \sum_{ik} V^{ik}_{g_{1}} \ket{i}\bra{k} \sum_{lj} V^{lj}_{g_{2}} \ket{l}\bra{j}
=\sum_{ikj} V^{ik}_{g_{1}}  V^{kj}_{g_{2}} \ket{i}\bra{j} \nn
&= \sum_{ij}  b(g_{1},g_{2}) V^{ij}_{g_{1}g_{2}} \ket{i}\bra{j}= b(g_{1} ,g_{2}) \ket{g_{1}g_{2} }
 \end{align}}
to obtain
\begin{align}
\mu\equiv \mathtikz{\muA{0}{0} }  : \ket{g_{1}} \otimes  \ket{g_{2}} = b(g_{1} ,g_{2}) \ket{g_{1}g_{2} },
\end{align} 
which is the same as the closed multiplication \eqref{fusion} and the multiplication rule of the twisted group algebra $\mathbb{C}_{b}[G]$.  In this case, the map \eqref{clopen} is also one to one, so the closed and open algebras are isomorphic.  This is a distinguishing feature of the equivariant TQFT, which sets it apart from an ordinary TQFT where the closed algebra is always commutative while the open algebra may not be.  
\paragraph{Summary of extended TQFT and quantum circuits}
For each 2D closed G-equivariant TQFT associated to a projective representation $\mathcal{V} \in H^{2}(G, U(1))$, the extension of the TQFT  introduces an open algebra $\mathcal{H}_{\text{open}}=\mathcal{V^*} \otimes \mathcal{V} $ corresponding to the Hilbert spaces on an interval.   Each state on $\mathcal{V^*} \otimes \mathcal{V} $ is associated to operators $A_{g}$ which can be interpreted as a quantum circuit  acting an ``edge mode"  qubit in $\mathcal{V}$. (In the group basis $A_{g}$ is just the projective representation matrix  $V_{g}$ associated to $\mathcal{V}$). When $G= \mathbb{Z}_{2} \otimes  \mathbb{Z}_{2}$, this circuit corresponds to the standard teleportation protocol. For general G, this defines a more general notion of teleportation by symmetry.

In the extended TQFT language, the identification of holonomies around the circle with a quantum circuit is expressed by the  closed to open map \eqref{clopen}.  This is an isomorphism which maps flux eigenstates on $\mathcal{H}_{S^1}$ to quantum circuits $A_{g}$, which implement the parallel transport of TQFT edge modes along an interval.  Later we will identify this notion of parallel transport with the entanglement holonomy defined in \cite{Czech:2018kvg}


\section{Towards a local description of MBQC}
\label{sec:local}
In the section \ref{sec:global},  we formulated a global version of MBQC on a 1D circle using the language of TQFT.     To relate this to the conventional, local description of MBQC in terms of single -qubit measurements, we need a procedure that maps the  states of the TQFT into states on a lattice of qubits.     This discretization procedure is really an encoding map that embedds the TQFT Hilbert space into a code subspace on the lattice. This encoding is implemented by repeated applications of the co product in the open algebra.  This coproduct defines a factorization map into an extended Hilbert space, which we will identify with the Hilbert space of the lattice model.     The factorization map introduces \emph{entanglement} edge modes, which are virtual particles that corresponding to the contracted edges of the MPS tensors $A_{g}$.  These edge modes cannot be accessed directly without cutting open a spatial slice.  Nevertheless, their entanglement holonomy can be extracted from the correlations among measurements  on a circular array of qubits. and constitutes the computation output  of MBQC in 1D.   We will argue that this provides an operational meaning to entanglement edge modes.  
\subsection{Factorization via the co-product}
The open algebra $\mathcal{H}_{\text{open}}= \mathcal{V^*} \otimes \mathcal{V} $ has both a product and a co-product.  The co-product is given by 
\begin{align} \label{cop}
 \Delta: \mathcal{H}_{\text{open}} &\to \mathcal{H}_{\text{open}}\otimes  \mathcal{H}_{\text{open}} \nn
\ket{g} &\to \frac{1}{|G|}\sum_{g_{1},g_{2}} b^{-1}(g_{1},g_{2}) \delta(g_{1}g_{2} ,g) \ket{g_{1}} \ket{g_{2}} ,
\end{align} 
 As emphasized in \cite{Wong:2022eiu},  $\Delta$ can be interpreted as a factorization map in to an extended Hilbert space  defined via the  path integral evolution:
\begin{align}
\Delta=\mathtikz{\deltaA{0}{0} ;\node  at (0,-1) {$e$}}
\end{align} 
This euclidean process splits a single interval into two intervals separated by a ``stretched" entangling surface, given by the small semicircular arc labeled by $e$.   This denotes an entanglement boundary condition which is required to satisfy the shrinkability condition  \footnote{This boundary condition need not be unique, although it will be so in the examples we consider.}
\begin{align}
\mu \circ \Delta= \mathtikz{\deltaA{0}{0} ;\muA{0cm}{-1cm} ;\node  at (0,-1) {$e$} } = \mathtikz{\idA{0}{0} } = \text{Identity Map}.
\end{align} 
This condition ensures that the entangling surface is completely transparent, and does not alter the correlation functions of the original, unfactorized states.
More generally,  any hole surrounded by the boundary condition $e$ is required to be shrinkable. \footnote{ In a 2D TQFT, the shrinkability constraint can be captured by a single diagramatic relation \cite{Donnelly:2018ppr}: \begin{equation} \label{ebraneaxiom}
 \qquad \mathtikz{\etaC{0cm}{0cm} } = \mathtikz{ \etaA{0cm}{0cm} \cozipper{0cm}{0cm}
\draw (0cm,0.5cm) node {\footnotesize $e$};
\draw (0cm,-0.25cm) node {\footnotesize $e$}},
\end{equation} which implies all holes created by an entanglement boundary can be closed.}.  In quantum information language, the shrinkability condition says that the factorization map is an isometry which encodes the Hilbert space of one interval into a code subspace inside the Hilbert space of two interval.  

In general,  the shrinkability constraint requires that the extended Hilbert space contains a suitable set of  \emph{entanglement} edge modes \cite{Donnelly:2018ppr}.  These entanglement edge modes are manifest in representation basis for $ \mathcal{H}_{\text{open}} \otimes \mathcal{H}_{\text{open}}$
\begin{align}
\ket{ V \,\bar{i}\,j} \otimes \ket{ V\, \bar{k}\,,l},\qquad i,j,k,l= 1,\cdots \dim V ,
\end{align} 
where the states labeled by $j$ and $k$ are the ``extra" edge mode degrees of freedom introduced by the entangling surface.   Unlike the states at the physical boundary labelled by $i$ and $l$, the entanglement edge modes are always entangled under the embedding defined by $\Delta$.  We can see this by writing action of the co-product  on the MPS wavefunctions $V_{il}(g)$ 
\begin{align}
\Delta : \sum_{g} V_{il}(g) \ket{g} \longrightarrow \sum_{g_{1},g_{2}}  \sum_{j} b(g_{1},g_{2})^{-1} V_{ij}(g_{1} ) V_{jl}(g_{2} ) \ket{g_{1}} \ket{g_{2}},
\end{align}
in which the entanglement edge mode index j is contracted.     In the MPS descriptions, these correspond to degrees of freedom on the contracted virtual edges, which are not directly accessible by local measurements.
\begin{align}
 \includegraphics[scale=.4]{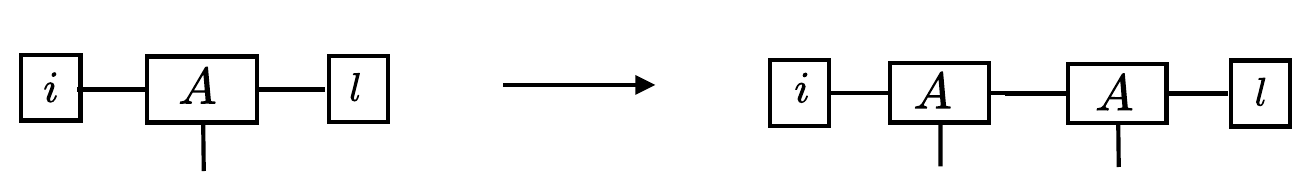}
\end{align}    

We can iterate this procedure to split the interval repeatedly: 
\begin{align}\label{seq}
 \mathcal{H}_{\text{open}}  \overset{\Delta}{\longrightarrow} \mathcal{H}_{\text{open}} \otimes \mathcal{H}_{\text{open}} \overset{1\otimes \Delta}{\longrightarrow} \mathcal{H}_{\text{open}} \otimes \mathcal{H}_{\text{open}} \otimes \mathcal{H}_{\text{open}} \longrightarrow \cdots  
 \end{align} 

After $N$ applications of $\Delta$, the original Hilbert space $\mathcal{H}_{\text{open}} $ is embedded into an extended Hilbert space $\mathcal{H}_{\text{open}}^{\otimes N+1} $, which we associate with the Hilbert on a lattice where  each $\mathcal{H}_{\text{open}} $ factor describes a local Hilbert space.   This procedure encodes the TQFT states into entangled lattice states that belong to the fixed point of a 1D SPT phase with symmetry $G$. 
\paragraph{Cluster state on a chain}
As an example of this encoding of TQFT states,   consider applying the sequence of embeddings \eqref{seq} to the states $\ket{A_{ij}} $ defined in \eqref{Aij}.   This produces entangled states on a 1D lattice
\begin{align}
\ket{A_{ij}} &\to  \ket{A_{ij}}_{\text{lat}} \equiv \sum_{g_{1},\cdots g_{N} \in \mathbb{Z}_{2} \otimes \mathbb{Z}_{2} }  b(g_{N},g_{N-1})^{-1} b(g_{N-3},g_{N-4})^{-1} \cdots b(g_{2},g_{1})^{-1}   (A_{g_{N} g_{N-1} \cdots g_{1}})_{ij} \ket{g_{N}} \otimes \ket{g_{2}} \cdots \ket{g_{1}}  \nn
&=\sum_{g_{1},\cdots g_{N} \in \mathbb{Z}_{2} \otimes \mathbb{Z}_{2} }  b(g_{N},g_{N-1})^{-2} b(g_{N-3},g_{N-4})^{-2} \cdots b(g_{2},g_{1})^{-2}  (A_{g_{N}} A_{g_{N-1}} \cdots A_{g_{1}})_{ij} \ket{g_{N}} \otimes \ket{g_{N-1}} \cdots \ket{g_{1}} \nn
\end{align} 

\begin{figure}[h]
\centering
\includegraphics[scale=.25]{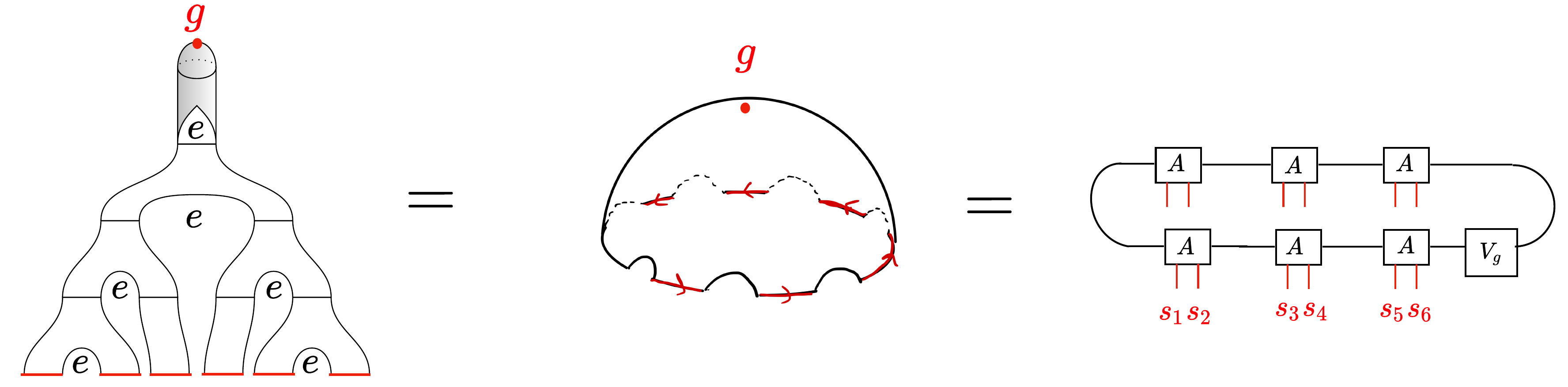}
\caption{A discretization of a quantum state $\ket{1}$ on the circle can be obtained by first applying the closed to open map, and then repeatedly applying the co-product. }
\label{discrete}
\end{figure}

When $G= \mathbb{Z}_{2}\otimes  \mathbb{Z}_{2} $, the phases square to 1\footnote{For general G, these phases do not affect the implemenation of MBQC on these resource states}, so we just have
\begin{align}\label{MPScluster}
\ket{A_{ij}}_{\text{lat}} = \sum_{g_{1},\cdots g_{N} \in \mathbb{Z}_{2} \otimes \mathbb{Z}_{2} }   (A_{g_{N}} A_{g_{N-1}} \cdots A_{g_{1}})_{ij} \ket{g_{N}} \otimes \ket{g_{N-1}} \cdots \ket{g_{1}},\qquad i,j=1,2.
\end{align}
We will refer to these state as \emph{cluster states} on a chain\footnote{These states satisfy more general boundary conditions than the usual definition of the cluster state, but we retain the same nomenclature for convenience.}. 
\paragraph{Cluster state on a circle} 
In a similar fashion, we discretize TQFT states on a circle by first applying the closed to open map on the twisted state $\ket{V_{g}} \in \mathcal{H}_{S^{1}}$ followed by repeated iterations of $\Delta$.  The resulting cobordism is shown in figure \ref{discrete}, which describes the embedding of  $\ket{V_{g}}$ into the extended Hilbert space on  the circle.    The image of  $\ket{V_{g}}$ is the matrix product states
\begin{align}\label{circle}
\ket{V_{g}} \to \ket{V_{g}}_{\text{lat}}= \sum_{h_{1},\cdots h_{N} \in \mathbb{Z}_{2} \otimes \mathbb{Z}_{2} } \tr( V_{g} A_{h_{N}} A_{h_{N-1}} \cdots A_{h_{1}}) \ket{h_{N}} \otimes \ket{h_{N-1}} \cdots \ket{h_{1}} 
\end{align} 
We emphasize again that the lattice Hilbert space does not contain the entanglement edge modes which live on the contracted virtual edges: nevertheless these virtual degrees of freedom play an important role as the logical qubit in 
MBQC. 
\subsection{The fully local description of the cluster state}
\label{sec:reference}
When $G= \mathbb{Z}_{2} \otimes \mathbb{Z}_{2} $,  the basis states $\ket{g}$ on each lattice site can be identified with the  entangled bell pairs $\ket{\beta_{s_{1}s_{2}}}$ which appeared in the teleportation protocol in figure \ref{tel}.  
However, to implement MBQC, we need to obtain a fully local description of the cluster state in which these bell pairs are replaced by un-entangled product states. In terms of the teleportation protocol, this corresponds to replacing bell measurements with single qubit measurements.  This is easily accomplished by observing that the bell pairs  can be created from a product state using a Cphase and Hadamard gate. 
In a basis $\ket{s_{i}}$ of $X$ eigenstates satisfying  
 \begin{align}
      X \ket{s_{i}} &= (-1)^{s_{i}} \ket{s_{i}} , \qquad s_{i}=0,1
 \end{align}
the Cphase gate is defined by
\begin{align}
\includegraphics[scale=.3]{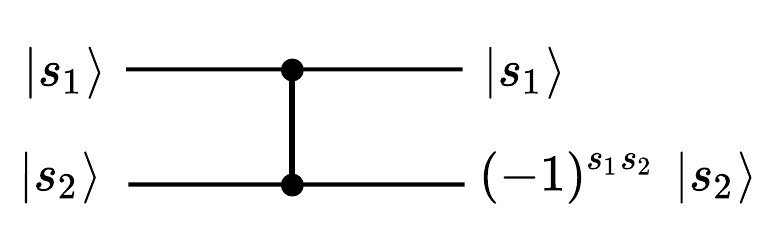}
\end{align} 
while the Hadamard gate is defined by
\begin{align}
    H=\frac{1}{\sqrt{2}} (X+Z)
\end{align}

The bell pair preparation and bell measurements in the teleportation circuit in figure \ref{tel} can then be replaced using the identities
\begin{align}
\includegraphics[scale=.3]{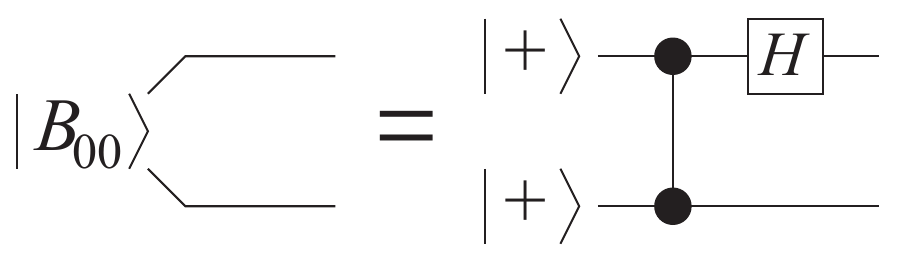}\qquad\qquad
\includegraphics[scale=.3]{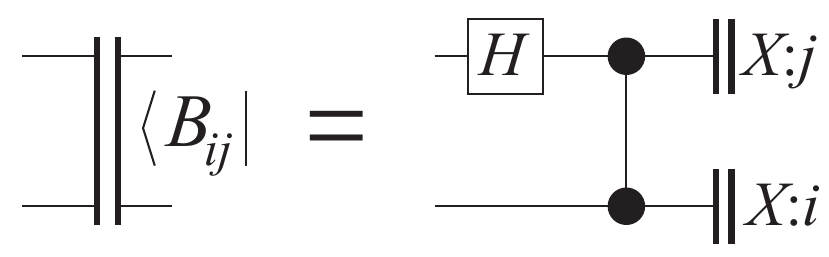}
\end{align}
The Hadamard gates will cancel, leaving behind a sequence of cphase gates that defines the lattice MPS tensor:
\begin{align}
\label{MPST}
\includegraphics[scale=.4]{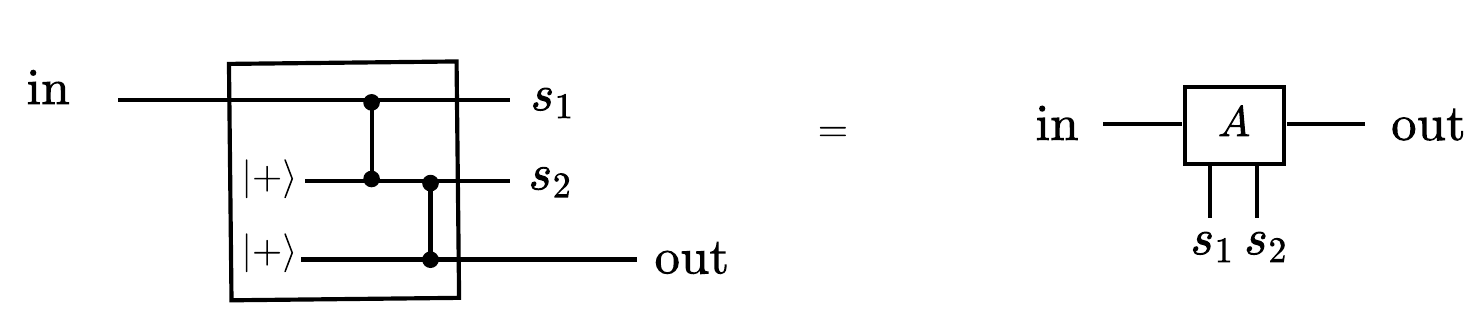}
\end{align} 
To identify this circuit with the abstract MPS tensor \eqref{MPS}, we 
 need a consistent assignment of group labels to the product state $ \ket{s_{1}} \otimes \ket{s_{2}}  $.   This can be achieved via a lattice version of the projective representation $l(h)$ defined in \eqref{lh} :
\begin{align}\label{l(ij)}
\tilde{l}(00)&= 1\otimes 1,\quad \tilde{l}(01) = X_2 \otimes Z_1, 
\quad \tilde{l}(10) = Z_2 \otimes 1 ,\quad \tilde{l}(11)= Z_2 X_2 \otimes Z_1
\end{align} 
Given a reference state $\ket{0}\ket{0}$ on the lattice, we can use this representation to assign group labels on the lattice: 
\begin{align}\label{glat}
\ket{g}_{\text{lat}} \equiv \tilde{l}(g) \ket{0}\ket{0}
\end{align} 

This defines a lattice realization of the $b$ regular representation\footnote{ We could have defined $\tilde{l}$ by just imposing \eqref{lh} on the lattice. But we want to connect to the representation $\tilde{l}$ that is used in the SPT literature.  In order for the action of $\tilde{l}$ to match with the regular representation, we need to define  $\ket{g}_{\text{lat}}$ as in \eqref{glat}.   }: \eqref{lh}
\begin{align}
    \tilde{l}(h) \ket{g}_{\text{lat}}  &\equiv \tilde{l}(h)  \tilde{l}(g)\ket{0}\ket{0}\nn
    &= b(h,g) \tilde{l}(hg) \ket{0}\ket{0} \nn
    &= b(h,g) \ket{hg}_{\text{lat}} 
\end{align}
Since this representation is isomorphic to \eqref{lh}, we will henceforth drop the tilde and the lat. subscripts.  However it is important to remember that the mapping of the abstract intertwiner $A_{g}$ to the lattice MPS tensor involves a choice of reference state.

Expressing the MPS tensors in the cluster state \eqref{MPScluster}  in terms of the quantum circuit in \eqref{MPST} makes it manifest that the cluster state can be created from a product state with a finite depth\footnote{Here finite depth refers to the fact that the number of layers in the unitary is independent of the system size} unitary.    This can be achieved by simply re-ordering the sequence of (commuting) cphase gates into two local Unitaries: 
\begin{align}
\includegraphics[scale=.3]{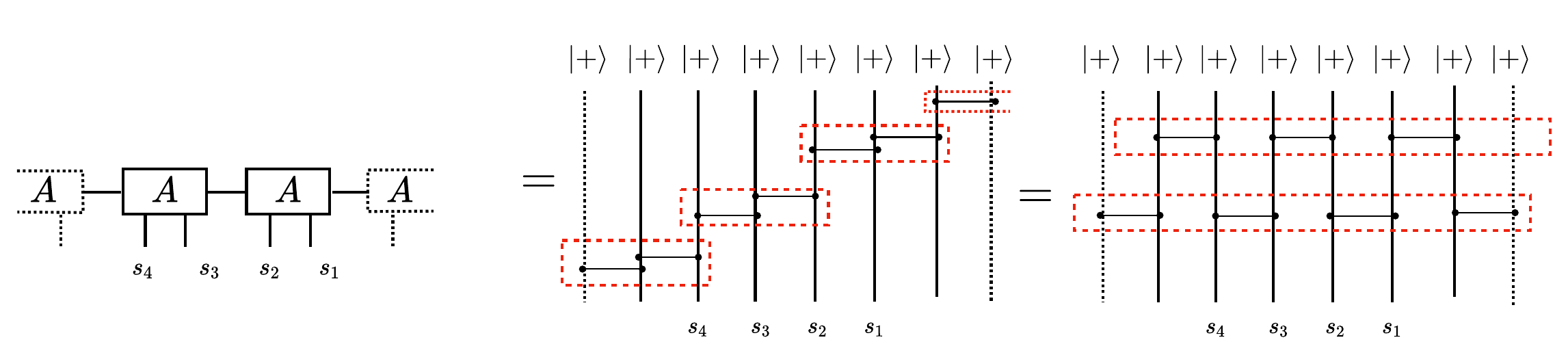}
\end{align} 
The fact that such a finite depth creation process exist is a general property of an SPT phase.  Thus, we can take a more general perspective in which the lattice serves as a blank canvas on which we can engineer any G-equivariant TQFT, where $G$ characterizes the entanglement structure of states belonging to a code subspace.
\subsubsection{Hilbert space and flux sectors on the lattice} 
The discretization process described above maps the closed TQFT Hilbert space $\mathcal{H}_{S^1}$ onto a Hilbert space $\mathcal{H}_{S^{1}}^{\text{lat}}$ on a periodic lattice of $N$ qubits.   $\mathcal{H}_{S^{1}}^{\text{lat}}$ can be  organized into flux sectors in a manner that parallels the decomposition of the closed TQFT Hilbert space \eqref{HS}:
\begin{align} \label{Hglat}
\mathcal{H}_{S^{1}}^{\text{lattice}} &= \oplus_{g} \mathcal{H}_{g}^{ \text{lattice}},\qquad g\in G\nn
\mathcal{H}_{g}^{ \text{lattice}} &\equiv   \text{span} \{\ket{h_{1}} \otimes \ket{h_{2}} \cdots \ket{h_{N}} ; \, \prod_{i}h_{i} =  g \} 
\end{align} 
On the lattice, the flux labels $g$ are measured by the charges that generate the global symmetry of the SPT phase. 

For example, when  $G= \mathbb{Z}_{2} \otimes \mathbb{Z}_{2}$, these charges are given by
\begin{align}\label{Q}
   (Q_{1},Q_{2}), \qquad 
Q_{1} = \prod_{a =1}^{N-1} X_{2a-1} ,\qquad 
Q_{2} = \prod_{a=1 }^{2N} X_{2a} 
\end{align} 
 These charges are the image under $\Delta$ of the symmetry operation $\alpha_{g} $ defined in \eqref{alpha}, which forms the symmetry data of the closed $\mathbb{Z}_{2} \otimes \mathbb{Z}_{2}$-equivariant TQFT.  Since the $X$ basis $\ket{h_{i}}$ diagonalizes this symmetry, it is referred to as the \emph{symmetry} basis in this context. 

In its usual lattice formulation, the cluster state is viewed as the ground state of the cluster Hamiltonian
\begin{align}
    H = \sum_{a}  Z_{a-1}X_{a} Z_{a+1}
\end{align}
which consists of a sum of stabilizers $Z_{a-1}X_{a} Z_{a+1}$.   The charges $Q_{1}$ and $Q_{2}$ commute with $H$ and therefore define a symmetry of the system.  The sectors $\mathcal{H}_{g}^{ \text{lattice}}$ do not mix under operators that commute with the symmetry; it is in this sense that they form ``superselection" sectors.   However, from the point of view of the lattice, there are no superselection rules between flux sectors.  Indeed we will see that 
MBQC implements quantum computations by manipulating the definition of these flux sectors. 

\section{Gauge symmetry from entanglement}
\label{sec:gaugesym}
 In section \ref{sec:global}, we motivated G-equivariant TQFT as the theory which arises upon coupling a matter TQFT to a background gauge field. However, the abstract formulation of this TQFT does not make reference to any local gauge symmetries.  This formalism circumvents the  usual gauge redundant, local degrees of freedom by working directly with the holonomy variables that label the twisted states $\ket{V_{g}}$ .  Formally, G-equivariant TQFT is just a Turaev algebra that describes the fusion of fluxes. 
 
On the other hand, once we have mapped the TQFT Hilbert space into a code subspace on the lattice, we can define local gauge transformations on the extended Hilbert space that leaves the twisted states $\ket{V_{g}}_{\text{lat}} $ invariant.   These are generated by stabilizers that characterize the entanglement structure of  $\ket{V_{g}}_{\text{lat}}$. In addition to the Hilbert space definition, we will give an operational interpretation of this gauge symmetry as transformations that preserve the circuit simulation implemented by MBQC.  From this definition, it will be manifest that in MBQC, the entanglement holonomies defined in \cite{Czech:2018kvg} are exactly the holonomies of a G-equivariant TQFT.



\subsection{Gauge symmetry and stabilizer transformations}
In our extended TQFT formulation of MBQC  a gauge symmetry arises because the encoding map that embedds TQFT states into the extended Hilbert space is not surjective.  This is because the co-product $\Delta$ always maps onto a subspace invariant under a combination of left/right edge mode symmetries.   
If we denote the co product acting on the $ith$ block by 
\begin{align}
\Delta^{i}  : \mathcal{H}_{\text{open}}^{i}   \to \mathcal{H}_{\text{open}}^{i}  \otimes \mathcal{H}_{\text{open}}^{i+1} 
\end{align} 
then this invariance is characterized by
\begin{align} \label{inv}
       l(h)_{i+1} r(h_{i}) \Delta^{i} = \Delta^{i} 
\end{align} 

This implies that the discretized TQFT states satisfy the constraints
\begin{align}\label{lr}
l(h)_{i+1} r(h)_{i} \ket{V_{g}}_{\text{lat}}=\ket{V_{g}}_{\text{lat}},  \qquad l(h)_{i+1} r(h)_{i} \ket{A_{IJ}}_{\text{lat}}=\ket{A_{IJ}}_{\text{lat}}
\end{align}
in the bulk of the chain.  These constraints can be viewed as gauge constraints that generate \emph{local} gauge transformations in the extended Hilbert space.  They act on  $ \mathcal{H}_{\text{open}}^{i}  \otimes \mathcal{H}_{\text{open}}^{i+1}$ according to 
 \begin{align}\label{gauge}
l(h)_{i+1}r(h)_{i} \ket{g_{i}}  \ket{g_{i+1} } =  b(h,g_{i+1}) b(g_{i},h^{-1}) \ket{g_{i}h^{-1}}  \ket{h g_{i+1} } ,
 \end{align} 

Note that the phases in \eqref{gauge} are important because they make this local representation of $G$ projective.   To be explicit, lets consider $G=\mathbb{Z}_{2} \otimes \mathbb{Z}_{2} $.    Writing $l(h)$ and $r(h)$ in terms of single qubit Pauli operators shows that \eqref{gauge} are the stabilizer constraints of the cluster state. 
\begin{align}\label{stab}
Z_{a-1}X_{a}Z_{a+1}\ket{V_{g}}_{\text{lat}} =\ket{V_{g}}_{\text{lat}}
\end{align} 
The presence of the $X$ operator in the stabilizer is what makes these gauge transformations projective, and  distinguishes them from the usual gauge transformations of lattice gauge theory.  To highlight the difference with lattice gauge theory, we will refer to the transformations \eqref{gauge} as MBQC gauge symmetries. 
\subsection{Entanglement holonomy and deterministic MBQC }

To understand the operational aspects of the MBQC gauge symmetry, it is useful to begin by formulating deterministic MBQC in the MPS language \cite{PhysRevA.76.052315} . In our extended TQFT formalism, deterministic MBQC refers to a protocol for simulating a unitary gate $V_{g}$  using a twisted state on the lattice:
\begin{align}\label{circle}
\ket{V_{g}}_{\text{lat}}= \sum_{h_{1},\cdots h_{N} \in \mathbb{Z}_{2} \otimes \mathbb{Z}_{2} } \tr( V_{g} A_{h_{N}} A_{h_{N-1}} \cdots A_{h_{1}}) \ket{h_{N}} \otimes \ket{h_{N-1}} \cdots \ket{h_{1}} 
\end{align} 
In deterministic MBQC, we measure each lattice site ( or pairs of lattice sites in the case of the cluster state) in the symmetry basis $\ket{h_{i}}$, moving in sequential order down the chain.  As we explained below equation \eqref{Ag},  a measurement of a single block with outcome $h_{i}$ produces a teleportation of the edge modes  in $\mathcal{V}$ given by the linear map $A_{h_{i}}$.  
\begin{figure}
    \centering
    \includegraphics[scale=.3]{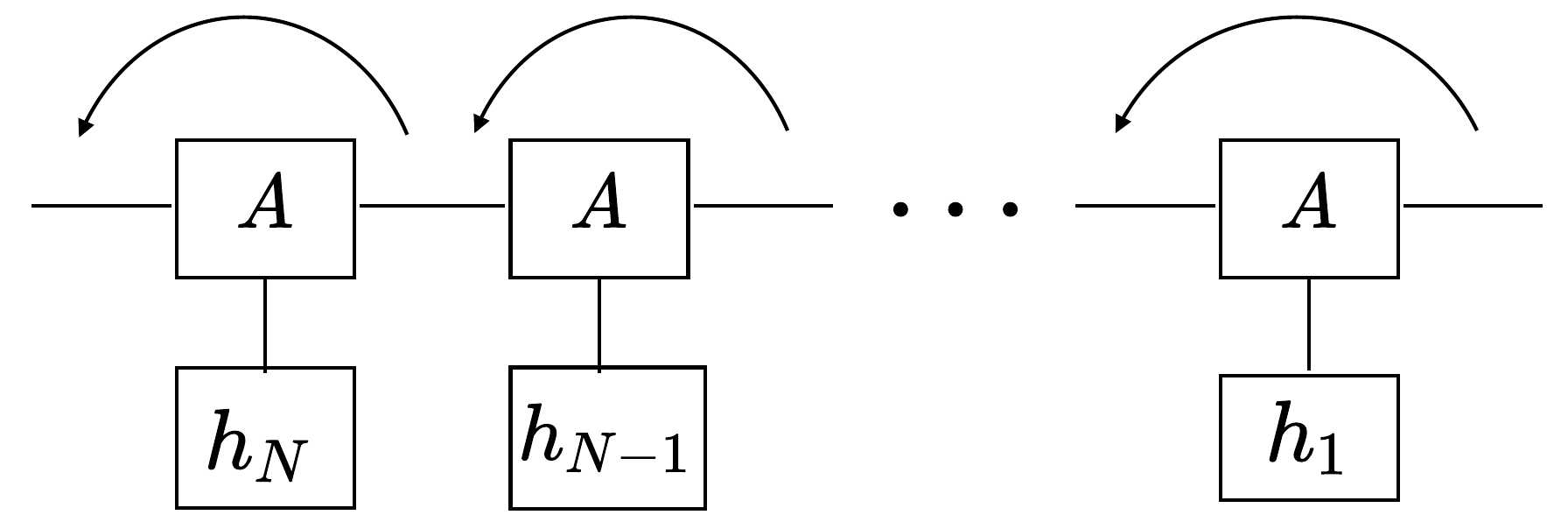}
    \caption{Each local measurement of the MPS describes the local parallel transport of a qubit between the edge Hilbert spaces}
    \label{fig:transport}
\end{figure}
According to \cite{Czech:2018kvg}, $A_{h_{i}} $ should be viewed as a local parallel transport of  the edge modes.  More precisely, we identify $h_{i}$ with a gauge connection on a principal $G$ bundle over a circle, and $A_{h_{i}}$ describes its action in the projective representation of $G$.  Since each measurement outcome is random, this local parallel transport is also random.   Due to this randomness,  no information can be extracted from  local measurements in any subregion of the circle.

To extract information from measurements of the twisted states $\ket{V_{g}}_{\text{lat}}$, we observe that they satisfy a global constraint which fixes the total parallel transport around the circle to be $V_{g}$ - after all these are the fluxes they were meant to encode !  The constraint is manifest in the MPS wavefunction 
\begin{align}
\tr( V_{g} A_{h_{1}} A_{h_{2}} \cdots A_{h_{N}}) = (\text{phase}) \tr(V_{g} A_{ h_{1}\cdots h_{N}}),
\end{align} 
which vanishes unless
\begin{align} \label{constraint}
g= \prod_{i} h_{i}. 
\end{align}
This implies that  even though the local parallel transport $A_{h_{i}}$  is random, the total parallel transport experienced by the logical qubit is:
\begin{align}\label{VgA}
\prod_{i} A_{h_{i}}=\pm V_{g}
\end{align}
Since there are no boundaries in the spatial geometry, this logical qubit has to annihilate itself upon being teleported in a closed loop: for a circuit model description of this process, we refer the reader to  appendix \ref{sec:telecircle}.   Thus the entanglement holonomy associated to measurements along the symmetry basis coincides with the ``physical" holonomy assigned by the equivariant TQFT.   
Notice that the entanglement holonomy \eqref{VgA} is invariant under the action of the stabilizer transformations \eqref{gauge}.  This is obvious in the MPS notation when we apply the equivariant transformaiton laws \eqref{rg} and \eqref{lg}, which pushes the action of the gauge transformations \eqref{gauge} in to the virtual space, where they correspond to acting with $V_{h}V^{\dagger}_{h}=1 $ on a virtual edge.
Since the entanglement holonomy is exactly the MBQC circuit simulation, these are also the transformations that preserve MBQC.   

The constraint \eqref{constraint} has an important consequence for the processing of information. Even though the individual measurement outcomes are completely random,  information is encoded nonlocally in the total set of outcomes around the whole circle.  For example, when $G= \mathbb{Z}_{2} \otimes \mathbb{Z}_{2} $, there are two classibal bits of information encoded in the quantity
 \begin{align} \label{hol}
 g= ( \sum_{a \,\text{odd}} s_{a} , \sum_{a\, \text{ even} } s_{a} ) \in \mathbb{Z}_{2}\times \mathbb{Z}_{2},
 \end{align}  
 where $(-1)^{s_{a}}, a= 1,\cdots 2N$ labels the X eigenvalues of the qubits.  This is a non local observable measured by the charges $Q_{i}, i=1,2$ of equation \eqref{Q}, and labels the different flux sectors on the circle.  These fluxes are computed directly from the measurement outcomes and constitutes the computational output of deterministic MBQC.   Following \cite{Wong:2022mnv} we will refer to these fluxes as MBQC holonomies to distinguish them from the entanglement holonomies \eqref{VgA}.

Finally, note that the $X$ basis states corresponding to different measurement outcomes $ \{s_{a}\}_{ a= 1,\cdots 2 N} $ with the same computational output \eqref{hol} are related by the gauge transformations \eqref{gauge}.
This is because the stabilizers $K_{a}=Z_{a-1}X_{a}Z_{a+1}$ flip an even number of spins on either the odd or even sublattice, so their effect on the measurement record is:
\begin{align}
    K_{a}: \quad &s_{a-1} \to s_{a-1} +1\nn
    &s_{a} \to s_{a}  \nn 
    &s_{a+1} \to s_{a+1} +1 
\end{align}
Thus,  ``pure gauge" fluctuations capture the random, local degrees of freedom that contain no information: they are the fluctuations that remain within a single superselection sector denoted by $\mathcal{H}_{g}^{\text{lattice}}$ in \eqref{Hglat}.  On the other hand, the gauge invariant degrees of freedom carries quantum information that is non-locally encoded in the measurement outcomes.  This provides an information theoretical interpretation of MBQC gauge symmetry, which will generalize to the probabilistic MBQC discussed below. 

\section{Probabilistic MBQC and measurement adaptations} \label{sec:PMBQC}

In the previous section, we gave an operational definition of the  gauge symmetry of deterministic MBQC.  This involved measurements of a resource state in an SPT phase, using the symmetry basis.    Below we will give a gauge theory formulation of the general MBQC protocol in which the qubits are measured in an adaptive basis, which enables the simulation of a general unitary gate.   

\subsection{Programming by state preparation}
Following the discussion in section \ref{sec:1Dsim}, we begin by explaining how to produce an arbitrary single qubit unitary $U=\sum_{g} c_{g} V_{g}$ using the same measurement protocol defined for deterministic MBQC, but applied to a modified, U -dependent resource state.   This resource state arises via the encoding of the TQFT state  $\ket{U}= \sum_{g} c_{g} \ket{V_{g}} $ (introduced in \eqref{U}) onto the lattice:
\begin{align}\label{Uket1}
\ket{U}_{\text{lat}}&\equiv \sum_{g} c_{g} \ket{V_{g}}_{\text{lat} } \\
&= \sum_{h_{1},\cdots h_{N} \in \mathbb{Z}_{2} \otimes \mathbb{Z}_{2} } \tr((\sum_{g} c_{g}  V_{g} )A_{h_{1}} A_{h_{2}} \cdots A_{h_{N}}) \ket{h_{1}} \otimes \ket{h_{2}} \cdots \ket{h_{N}}\nn
&=\sum_{h_{1},\cdots h_{N} \in \mathbb{Z}_{2} \otimes \mathbb{Z}_{2} } \tr( U  A_{h_{1}} A_{h_{2}} \cdots A_{h_{N}}) \ket{h_{1}} \otimes \ket{h_{2}} \cdots \ket{h_{N}}  ,\label{Uket}
\end{align}

Since $\ket{U}_{\text{lat}}$ does not belong to a single flux sector as defined in \eqref{Hglat} ,  single qubit measurements in the symmetry basis $\ket{h_{i}}$ will no longer give a definite computational output \eqref{hol}.   Instead, each MBQC holonomy $g$ will occur with a probability distribution given by 
 \begin{align}
     \text{Prob}(g)&= | \bra{h_{1}} \otimes \bra{h_{2}} \cdots \braket{h_{N} |U}|^{2} \text{ for states satsifying }\prod_{i} h_{i} = g  \nn
     &= |c_{g}|^{2}   
 \end{align}
This defines the computational output, which is interpreted as simulation of 
\begin{align}
    U=\sum_{g} c_{g} V_{g} 
\end{align}

Notice that it is still true that the ``pure gauge" fluctuations within each flux sector  do not carry any information. However, in addition to these fluctuations,  the state $\ket{U}_{\text{lat}}$ contains fluctuations that mixes the different sectors, and these fluctuations have non trivial information content that determines the unitary $U$.

\subsection{Probabilistic MBQC and generalized MBQC fluxes}
As discussed in section \ref{sec:1Dsim}, rather than changing the resource state,  MBQC provides a novel way to extract the computational output $ |c_{g}|^{2}$ by measuring $\ket{1}_{\text{lat}}$ in a cleverly chosen product basis.   We saw previously that the closed TQFT analogue (\eqref{rotate})of this basis is obtained from fusion of $\ket{V_{g}}$ with $ \ket{U^{\dagger}}$, which gives the states $\ket{U^{\dagger } V_{g} }$.    
Via our embedding map from the TQFT Hilbert space to the lattice, we can define an analogue of this fusion between $\ket{U}_{lat}$ and $\ket{V_{g}}_{lat}$.   In the MPS description, this corresponds to just inserting an operator $U$ before closing the chain in the circle.  This  results in the U-rotated twisted states:
\begin{align} \label{UdagVg}
\ket{U^{\dagger } V_{g} }_{\text{lat}} \equiv \sum_{h_{1},\cdots h_{N}  } \tr( U^{\dagger} V_{g}   A_{h_{1}} A_{h_{2}} \cdots A_{h_{N}}) \ket{h_{1}} \otimes \ket{h_{2}} \cdots \ket{h_{N}} ,
\end{align} 
These form a $|G|$ dimensional basis of a U-dependent code subspace $\mathcal{H}_{\text{code}}(U)$ that contains the resource state $\ket{1}_{\text{lat}}$.  

Indeed we can check explicitly that
\begin{align}\label{1lat}
    \ket{1}_{\text{lat}} = \sum_{g} c_{g} (U) \ket{ U^{\dagger} V_{g} }_{\text{lat}},
\end{align}
which is the lattice analogue of \eqref{1}.   (Below we will suppress the dependence of $c_{g}$ on $U$)

To understand how this equation relates to MBQC , we must consider the following question : what single-qubit measurements of $\ket{1}_{\text{lat}}$  can be used to extract the probabilities $|c_{g}|^{2}$ for a general choice of $U$ ?   Notice that the states $\ket{U^{\dagger } V_{g} }_{\text{lat}} $ are as entangled as the cluster state itself, so they do not make a good choice for a simple measurement basis.   
 Also, measurements in the symmetry basis $\ket{h_{i}}$ will no longer work, since this will return the computation output $( \sum_{a \,\text{odd}} s_{a} , \sum_{a\, \text{ even} } s_{a} ) =(0,0)$ with probablity 1.   
 
The results of the seminal work \cite{PhysRevLett.86.5188} provides an answer to this question by defining a \emph{temporally ordered} measurement basis which depends on the choice of circuit $U$ being simulated.  Earlier in section \ref{sec:review}, we defined this basis for the case $G=\mathbb{Z}_{2}\otimes \mathbb{Z}_{2} $ via the observables $O\big((-)^{q_{i}}\alpha_{i}\big) $ (eq.  \eqref{orot} ) measured in the protocol.   
Explicitly, this measurement basis consists of temporally ordered spin states:
 \begin{align}\label{tpb}
     \ket{\vec{q}, \vec{\mathbf{s}}}=\ket{q_{1},\mathbf{s}_{1}} \ket{ q_{2},\mathbf{s}_{2}} \cdots \ket{q_{N},\mathbf{s}_{N}}
 \end{align}
 where we used the bold font $\mathbf{s}_{i}=0,1$ to denote the spin pointing along an angle $(-)^{q_{i}}\alpha_{i}$ in the XY plane, and $q_{i}$ are the adaptive parameters defined in \eqref{CPR1}.  The rotation of the measurement angles enables the simulation of a general SU(2) gate, while the adaptations are needed to mitigate the randomness of the outcomes. 
 \paragraph{Reference states and group labels}
To understand the relation between the temporally ordered basis $  \ket{\vec{q}, \vec{\mathbf{s}}}$ and the basis $\ket{U^{\dagger}V_{g}}$ generating the code subspace, it is useful to assign group labels to $  \ket{\vec{q}, \vec{\mathbf{s}}}$.   This was explained in \cite{Wong:2022mnv}, but since it is crucial in what follows, we will review the construction here.   
First let us consider the temporally ordered measurements of a single block of two qubits on a cluster state. Temporal order means that the measurement angle of the second qubit is flipped when ever the first qubit is measured to be in the state $\mathbf{s}_{1}=1$.    For a general pair of outcomes $(\mathbf{s}_{1},\mathbf{s}_{2})$, this choice of measurements induces a transport of the edge mode given by \cite{Raussendorf:2002zji}
\begin{align} \label{ABU}
    A_{\mathbf{s}_{1},\mathbf{s}_{2}} (\alpha_{1},\alpha_{2})  = B_{\mathbf{s}_{1},\mathbf{s}_{2}}, U(\alpha_{1},\alpha_{2}),
    \end{align}
    where $U$ captures the desired local circuit simulation
    \begin{align}
U(\alpha_{1},\alpha_{2})\equiv A_{00} (\alpha_{1},\alpha_{2}) =
\exp\left(-i\frac{\alpha_2}{2}X\right) \exp\left(-i\frac{\alpha_{1}}{2}Z\right)
\end{align} 
while the remaining operator is a random byproduct:
\begin{align} \label{by}
\qquad B_{\mathbf{s}_{1}\mathbf{s}_{2}} = X^{\mathbf{s}_{1}}Z^{\mathbf{s}_{2}}
\end{align}

We can label the outcomes 
$(\mathbf{s}_{1},\mathbf{s}_{2})$ with group elements because the lattice representation \eqref{glat} of $l(g)$  permutes these outcomes, as one can see from the commutation relations
\begin{align}
\label{XOX}
X_{i}\, O(\alpha_{i}) &= O(-\alpha_{i})\, X_{i}\nn
Z_{i}\, O(\alpha_{i}) &= -O(\alpha_{i})\, Z_{i}
\end{align}  
and the explicit representation of $l(g)$ in \eqref{l(ij)}.
In particular, the commutation relation of $O(\alpha_{i})$ with $X$ for non-zero $\alpha_{i}$ leads to the measurement adaptation of the second qubit.
Thus, given a  reference state $\ket{0}_{\alpha_{1}} \ket{0}_{\alpha_{2}}$, we can then define: 
\begin{align} \label{lgalpha}
\ket{\mathbf{h}} \equiv l(\mathbf{h}) \ket{0}_{\alpha_{1}} \ket{0}_{\alpha_{2}}
\end{align} 
This is the direct generalization of \eqref{glat}.
\paragraph{Circuit simulation in the temporally ordered basis}
In terms of the basis \eqref{lgalpha}, the local circuit simulation can be expressed in MPS notation as \cite{Else_2012}:
\begin{align}\label{localsym}
   A_{\mathbf{h}} =  \mathtikz{  \draw (0,0) rectangle (.7,.4);\draw (0,-.3) rectangle (.7,-.7);\draw (-.4,.2) -- (0,.2) ; \draw (.7,.2) -- (1.1,.2) ; \draw (.2,0)--(.2,-.3);\draw (.5,0)--(.5,-.3);\draw (0.33cm,0.20cm) node { $A$};\draw (.35,-.5) node { \footnotesize $\mathbf{h}$}} = \mathtikz{  \draw (0,0) rectangle (.7,.4);\draw (0,-.3) rectangle (.7,-.7);\draw (-.4,.2) -- (0,.2) ; \draw (.7,.2) -- (1.1,.2) ; \draw (.2,0)--(.2,-.3);\draw (.5,0)--(.5,-.3);\draw (0.33cm,0.20cm) node { $A$};\draw (.35,-.5) node { \footnotesize $\mathbf{1}$} ;  \draw (-.8,0) rectangle (.-.4,.4);\draw (-.6,.2) node { \footnotesize $B$};\draw (-1.2,.2) -- (-.8,.2) } 
   =\mathtikz{  \draw (0,0) rectangle (.7,.4);\draw (-.4,.2) -- (0,.2) ; \draw (.7,.2) -- (1.1,.2) ;\draw (0.33cm,0.20cm) node { $U$};\draw (-.8,0) rectangle (.-.4,.4);\draw (-.6,.2) node { \footnotesize $B$};\draw (-1.2,.2) -- (-.8,.2) } 
\end{align}
In the second equality we used \eqref{lgalpha} to write the measured state in terms of the action of $l(\mathbf{h}$ on the reference state $\ket{1}$, and then applied the identity \eqref{lg} to push this operation to the left edge, where it becomes the by product operator \eqref{by}.   This gives a symmetry derivation of equation \eqref{ABU}.

To obtain a general unitary we now have to combine multiple blocks together.   If we simply combine $N$ MPS tensor of the form \eqref{localsym},  the circuit simulation we take the form $$B_{N}U_{N}(\alpha_{2N},\alpha_{2N-1}) B_{N-1}U_{N-1}( \alpha_{2N-2},\alpha_{2N-3}) \cdots B_{1}U_{1}(\alpha_{2},\alpha_{1})$$ So to obtain the desired unitary we need to commute the by-product operators pass the $U's$ to put them on the left.  The propagation of these byproduct operators will flip the signs of the angles $\alpha_{2i},\alpha_{2i-1}$ that determine each local unitary. The adaptation of measurement basis is designed to compensate for this these random sign flips.  This is the origin of temporal order in MBQC, and it leads to a new feature in the choice of reference states:  the choice of reference in each block must also be adapted based on previous blocks in the chain. These temporally ordered reference states were defined in \cite{Wong:2022mnv} and reviewed in the appendix\footnote{As explained in \cite{Wong:2022mnv}, these reference states are a choice of section for the MBQC gauge bundle. The novelty is that the choice of section itself must change according to the measurement outcomes.  In this sense MBQC involves a quantum reference frame}. The upshot is that with respect to this reference state, we can assign group labels to the temporally ordered basis:
\begin{align} \label{hU}
\ket{\vec{q},\vec{\mathbf{s}}}=
\ket{\mathbf{h}_{N}}_{U} \ket{\mathbf{h}_{N-1}}_{U} \cdots \ket{\mathbf{h}_{1}}_{U}, \mathbf{h}_{i} \in  G,
\end{align}
where $\ket{\mathbf{h}_{i}}_{U}$ depends \footnote{notice that not just the group label, but the entire state at the ith block depends on the basis elements in the previous blocks. } on $\ket{\mathbf{h}_{j}}_{U},\,j<i$.
Here the subscript $U$ denotes the fact that the local measurement basis also depends on the circuit we want to simulate: for the case of $G= \mathbb{Z}_{2}\otimes \mathbb{Z}_{2}$ this dependence arise in the angles $\alpha_{i}$ on the $XY$ plane describing the measurement angles on each site.  

The essential property of temporally ordered basis \eqref{hU} is that it gives a change of basis that transforms the the MPS wavefunction  $A_{g_{N}}A_{g_{N-1}} \cdots A_{g_{1}}$ with respect to the symmetry basis into
\begin{align}\label{ABU}
\mathbf{A}_{\mathbf{h}_{N}}\mathbf{A}_{\mathbf{h}_{N-1}} \cdots \mathbf{A}_{\mathbf{h}_{1}}&= 
\sum_{g_{N},\cdots g_{1}} A_{g_{N}}A_{g_{N-1}} \cdots A_{g_{1}}  \bra{g_{N}} \cdots \bra{g_{1}} \ket{\mathbf{h}_{N}}_{U} \cdots \ket{\mathbf{h}_{1}}_{U} \nn&= B_{\mathbf{h}_{N}} \cdots B_{\mathbf{h}_{1}} U\nn 
\end{align}    
where  $\bra{g_{1}} \ket{\mathbf{h}_{N}}_{U} \cdots \ket{\mathbf{h}_{1}}_{U}$ is the change of basis matrix.   
This wavefunction describes a circuit simulation as follows. In a chain with open boundaries, an input qubit $\ket{\Psi_{0}}$ is prepared as an ``initial"  boundary condition for the MPS equation, so that measurements in the temporally ordered basis produces the edge state
\begin{align}
    \ket{\Psi_{2N+1}} = \mathbf{A}_{\mathbf{h}_{N}}\mathbf{A}_{\mathbf{h}_{N-1}} \cdots \mathbf{A}_{\mathbf{h}_{1}} \ket{\Psi_{0}}=B_{\mathbf{h}_{N}} \cdots B_{\mathbf{h}_{1}} U \ket{\Psi_{0}}
\end{align}
This is viewed as the implementation of a general unitary $U$ on $\ket{\Psi_{0}}$, up to the random by product $B_{\mathbf{h}_{N} \cdots \mathbf{h}_{1}}$. To compensate for the effects of the byproduct, one combines the measurement outcome $\mathbf{s}_{2N+1}$ of the final qubit $\ket{\Psi_{2N+1}}$ with the previous measurement outcomes into the computational output;
\begin{align}
    \mathbf{a}= \mathbf{s}_{2N+1} + \mathbf{s}_{2N-1} + \cdots \mathbf{s}_{1}
\end{align}
Measurement of this observable gives information about matrix elements of $U$ via the relation 
\begin{align}
\text{Prob}(\mathbf{a})= |\braket{\mathbf{a}
|U|\Psi_{0}}|^2
\end{align}
This is because measuring the final qubit probes the overlap
\begin{align}
    \braket{\mathbf{s}_{2N+1}| 
 \Psi_{2N=1}}= \braket{\mathbf{s}_{2N+1}| 
B_{\mathbf{h}_{N}} \cdots  B_{\mathbf{h}_{1}} U \ket{\Psi_{0}}}= \braket{\mathbf{a}
|U|\Psi_{0}},
\end{align}
where in the last equality we observed that the effect of the byproduct operators on $\ket{\mathbf{s}_{2N+1}}$ is to produce the state $\ket{\mathbf{a}}$(since each factor of Z in the byproduct flips the spin outcomes according to equation \eqref{XOX}).

Intuitively the string of random byproduct operators, act like a Wilson line and the final qubit acts like a charge that dresses this Wilson line to make a gauge invariant quantity.  To make this intuition precise, we will define a notion of MBQC holonomies and gauge transformations on the lattice.  This is particularly useful for MBQC on a circle where we tie ends of Matrix product state, in which case $U$ is measured indirectly through flux measurements as we explain below.  
\paragraph{Circuit Simulation with MBQC flux } 
In the temporarily ordered basis the MPS description of the cluster state is
\begin{align}\label{1temp}
    \ket{1}_{\text{lat}} &\equiv  \sum_{\mathbf{h}_{1},\cdots \mathbf{h}_{N} \in \mathbb{Z}_{2} \otimes \mathbb{Z}_{2}}
   \tr(  \mathbf{A}_{\mathbf{h}_{N}}\mathbf{A}_{\mathbf{h}_{N-1}} \cdots \mathbf{A}_{\mathbf{h}_{1}})\ket{\mathbf{h}_{N}}_{U} \otimes \ket{\mathbf{h}_{N-1}}_{U} \cdots \ket{\mathbf{h}_{1}}_{U}
    \nn
    &=\sum_{\mathbf{h}_{1},\cdots \mathbf{h}_{N} \in \mathbb{Z}_{2} \otimes \mathbb{Z}_{2} } \tr(   B_{\mathbf{h}_{N}} B_{\mathbf{h}_{N-1}} \cdots B_{\mathbf{h}_{1}} U) \ket{\mathbf{h}_{N}} \otimes \ket{\mathbf{h}_{N-1}} \cdots \ket{\mathbf{h}_{1}} 
\end{align}   
This expression is analogous to \eqref{Uket}, with the byproduct operators now playing the role of the MPS tensors: indeed our extremely compact notation was designed to highlight this similarity.   Here we would like to flesh out the analogy between by showing that  $\eqref{1temp}$ expresses the cluster state as a superposition of twisted states with definite \emph{MBQC} flux.    

Just as the symmetry basis can be divided into super-selection sectors corresponding to different $G$-flux, the temporally ordered basis \eqref{hU} also breaks up into superselection sectors. These are labelled by the MBQC flux defined by $g=\prod_{i}\mathbf{h}_{i} $ 
\begin{align} \label{Hglat}
\mathcal{H}_{S^{1}}^{\text{lattice}} &= \oplus_{g} \mathcal{H}_{g}^{ \text{lattice}}(U),\qquad g\in G\nn
\mathcal{H}_{g}^{ \text{lattice}}(U) &\equiv   \text{span} \{\ket{\mathbf{h}_{1}}_{U} \otimes \ket{\mathbf{h}_{2}}_{U} \cdots \ket{\mathbf{h}_{N}}_{U}; \, \prod_{i}\mathbf{h}_{i} =  g \} 
\end{align} 
The different sectors do not mix under operators that commute with the charge that measures the MBQC flux. 
As we show in appendix \ref{sec:stabgauge}, the temporally ordered basis in each flux sector form an orbit under the action of generalized stabilizers \eqref{gauge}. Moreover, as in deterministic MBQC, these stabilizers act as gauge transformations that preserve circuit simulation.   
Thus we see that the same lattice Hilbert space $\mathcal{H}_{S^{1}}^{\text{lattice}}$ admits different $U$-dependent decompositions into flux sectors.   
This means that when an MBQC technician rotates his measurement apparatus in a manner specified by $U$ and the temporal ordering, he is manipulating the flux decomposition of the Hilbert space.   

This ability to rotate the flux sectors is what allows MBQC to implement the different circuit simulation on the same resource state.  To see this go back to \eqref{1lat} and write the basis $\ket{ U^{\dagger} V_{g} }_{\text{lat}} $ that generate the code subspace in term of the temporally ordered basis. This is obtained by plugging in the change of basis matrix in \eqref{ABU} in to equation \eqref{UdagVg}
\begin{align}\label{udageV2}
\ket{ U^{\dagger} V_{g} }_{\text{lat}}= \sum_{\mathbf{h}_{1},\cdots \mathbf{h}_{N} \in \mathbb{Z}_{2} \otimes \mathbb{Z}_{2} } \tr( V_{g}  B_{\mathbf{h}_{1}} B_{\mathbf{h}_{2}} \cdots B_{\mathbf{h}_{N}}) \ket{\mathbf{h}_{1}}_{U} \otimes \ket{\mathbf{h}_{2}}_{U} \cdots \ket{\mathbf{h}_{N}}_{U}.
\end{align}
Here the matrices $B_{\mathbf{h}}$ are the same as $V_{\mathbf{h}}$ in  \eqref{UdagVg}, but we have changed notation to highlight the relation to the  byproducts in the MBQC protocol on a chain.   This way of writing the MPS wavefunction for $\ket{ U^{\dagger} V_{g} }_{\text{lat}}$ shows that they are states with definite entanglement holonomy $V_{g}$ when measured in the temporal basis: this means that doing MBQC on them simulates $V_{g}$ with probablity 1.    
We can now write the expansion of the cluster state in \eqref{1lat} as
\begin{align} \label{1rot}
  \ket{1}_{\text{lat}} = \sum_{g} c_{g} (U) \tr( V_{g}  B_{\mathbf{h}_{1}} B_{\mathbf{h}_{2}} \cdots B_{\mathbf{h}_{N}}) \ket{\mathbf{h}_{N}}_{U} \otimes \ket{\mathbf{h}_{N-1}}_{U} \cdots \ket{\mathbf{h}_{1}}_{U},
\end{align}
where we interpret the sum over $g$  as a sum over MBQC flux sectors on the lattice.  As described earlier, measuring this state in the basis $\ket{h_{i}}_{U}$ produces a simulation of $U$  acting on a virtual qubit via a sequence of teleportations.   However, instead of measuring this qubit directly, MBQC extracts information about $U$ by measuring the probability of obtaining an MBQC flux $g$:     
\begin{align}
   \text{prob} (g) \equiv \text{prob} (\prod_{i} \mathbf{h}_{i} ), \text{ such that } \prod_{i} \mathbf{h}_{i}  = g 
\end{align}
This in turn determines the coefficients $c_{g}$ which specifies U:
\begin{align}
    |c_{g}|^{2}(U) = \text{prob} (g) 
\end{align}
As we emphasized before, the programming of the unitary $U$ is achieved by changing the MBQC flux sectors.   Thus a fixed resource state can be used to achieve any circuit simulation.    Indeed, as discussed in section \ref{sec:global}, this works because the protocol for simulating $U$ involves the measurement of a \emph{relative} rotation between the flux of the resource state, and the flux of the twisted basis states.  So we can perform the same circuit simulation by either rotate the resource state as in \eqref{Uket} or rotate the definition of the fluxes themselves as in \eqref{1rot}. 


\section{Conclusion}
\paragraph{Recap}
In this work, we gave a formulation of  1+1 D MBQC using the framework of G-equivariant TQFT.   These TQFT's are known to be in one to one correspondence with 1+1 SPT phases, which are computational phases for MBQC \cite{Stephen:2016vgq}.  We explained how the flux Hilbert space of the G-equivariant TQFT plays the role of a code space for MBQC.   On the code space, the simulation of a unitary $U$ via  measurement adaptations is effected by fusing the flux eignenbasis with a state  $\ket{U^\dagger}$ specified by the simulation.  Thus, MBQC makes use of the fusion algebra on the code space, as determined by the TQFT, and not just the linear structure of the Hilbert space.  

To make the circuit simulation manifest in the TQFT, we introduced the \emph{extension} of the G-equivariant TQFT.   This produces a notion of circuit time evolution, corresponding to morphisms between boundary categories. This is the TQFT description of teleportation. 

The extended theory also provides a co- product which defines the encoding map from the TQFT to the lattice Hilbert space.  We explained how the properties of the co-product gives rise to the MBQC gauge symmetry which preserves the circuit simulation.  These are generated by the stabilizers.    The upshot is that our TQFT perspective shows how measurement adaptations effectively manipulate fluxes in a code space to produce a quantum computation. 

\paragraph{A information theoretical paradigm for gauge theory }

In a topological gauge theory, the  gauge invariant degrees of freedom are nonlocal: these are captured by holonomies. However, to couple the system to local probes, we need to introduce a local field $A_{\mu}(x)$, which can couple to a local current via the interaction $A_{\mu}(x)J^{\mu}(x)$.   The local description must have redundancies, and this is is captured by a gauge symmetry.   

 From this point of view, topological gauge theory is naturally suited to describe quantum computation, which involves the processing  of quantum information. Unlike classical information, quantum information can be non-locally supported on an entangled resource state.    This information is captured by a non local variable, which we refer to as an entanglement holonomy. MBQC is a protocol which seeks to extract these entanglement holonomies via  local measurements.    Doing so requires the introduction of gauge symmetries, which capture the redundancy in the local description of the entanglement holonomy.    These aspects of gauge theory are manifest in  MBQC on a the cluster ring: To extract two bits of information from local measurements of the  resource states, we must observe the correlations in the measurement outcomes along the entire ring.  We have shown that these two bits of information correspond to a bona fide  $\mathbb{Z}_{2}\otimes \mathbb{Z}_{2}$ holonomy of a gauge theory.   Moreover, different measurement outcomes which give the same holonomy are related by a gauge symmetry.   Thus the measurement outcomes play the role of the gauge field $A_{\mu}(x)$.  In this description, non gauge invariant observables are measurable, but they are random and contain no information.  This gives a concrete realization of the ideas in \cite{Rovelli_2014}.
\paragraph{Future directions}
In general, quantum information is composed of a mixture of local and non local elements \cite{Horodecki_2005}.   A general gauge theory of quantum information should capture both types of information.   It would be interesting to construct such a non-topological gauge theory to see if local gauge invariants will capture  the local components of quantum information.  
Another follow up to our work pertains to MBQC in higher dimensions, where universal computation can be realized.    As observed in \cite{Czech:2018kvg}, the higher dimensional protocol naturally involves junctions of Wilson lines.  We will report on this in future work. 

\section*{Acknowledgements}
It's a pleasure to thank Bartek Czech, Robert Raussendorf for initial collaboration on this project.   I was also like to thank Bartek Czech, Ho Tat Lam and Yale Fan for reading an early draft of this paper, and Kantoro Ohmori and Uri Kol for useful discussions.   GW is supported by the Science and Technology Facilities Council grant
ST/X000761/1.  GW was also supported by Harvard CMSA and the Oxford Mathematical Institute. 

\appendix

\appendix
\section{Super dense coding}
\label{superdense}
\begin{figure}[h] 
\centering
\includegraphics[scale=.3]{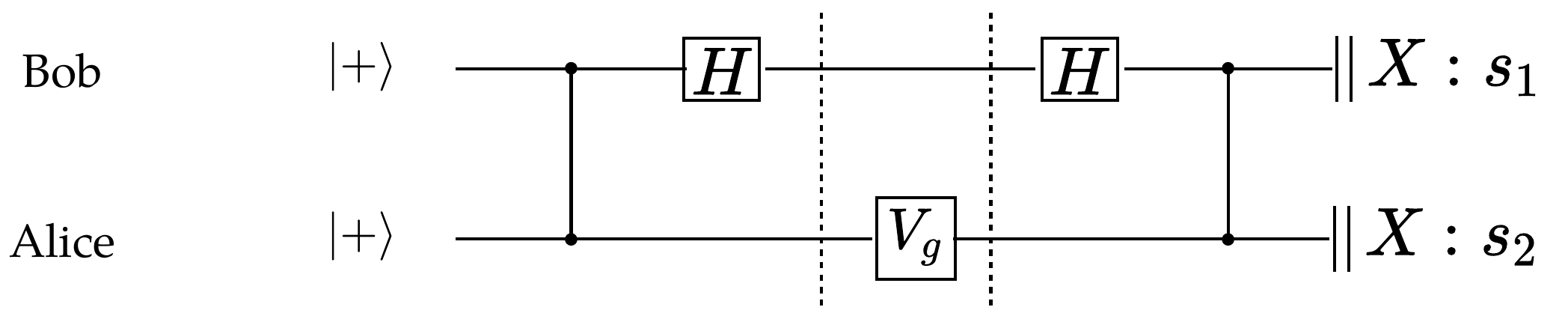}\qquad \
\caption{Circuit for superdense coding}
\label{dense}
\end{figure}

Superdense coding is a protocol which allows Alice to send to Bob a single qubit from which he can extract two classical bits of information. The circuit is shown in figure \ref{dense} 
\subsection{Generalization of super dense coding: teleportation in a closed loop}
\label{sec:telecircle}
\begin{figure}[h]
\includegraphics[scale=.2]{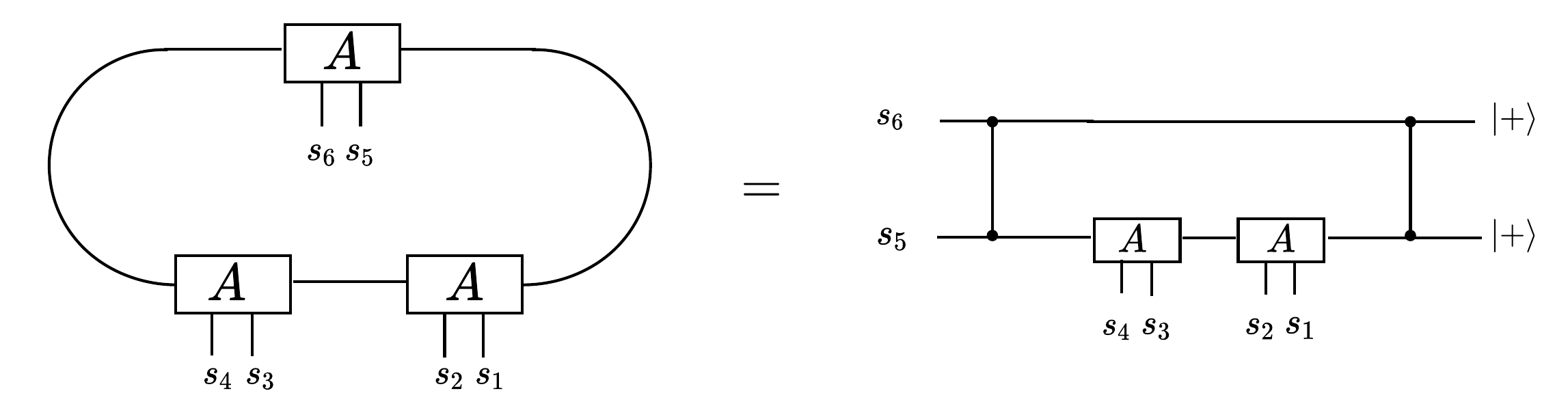} \qquad 
\includegraphics[scale=.2]{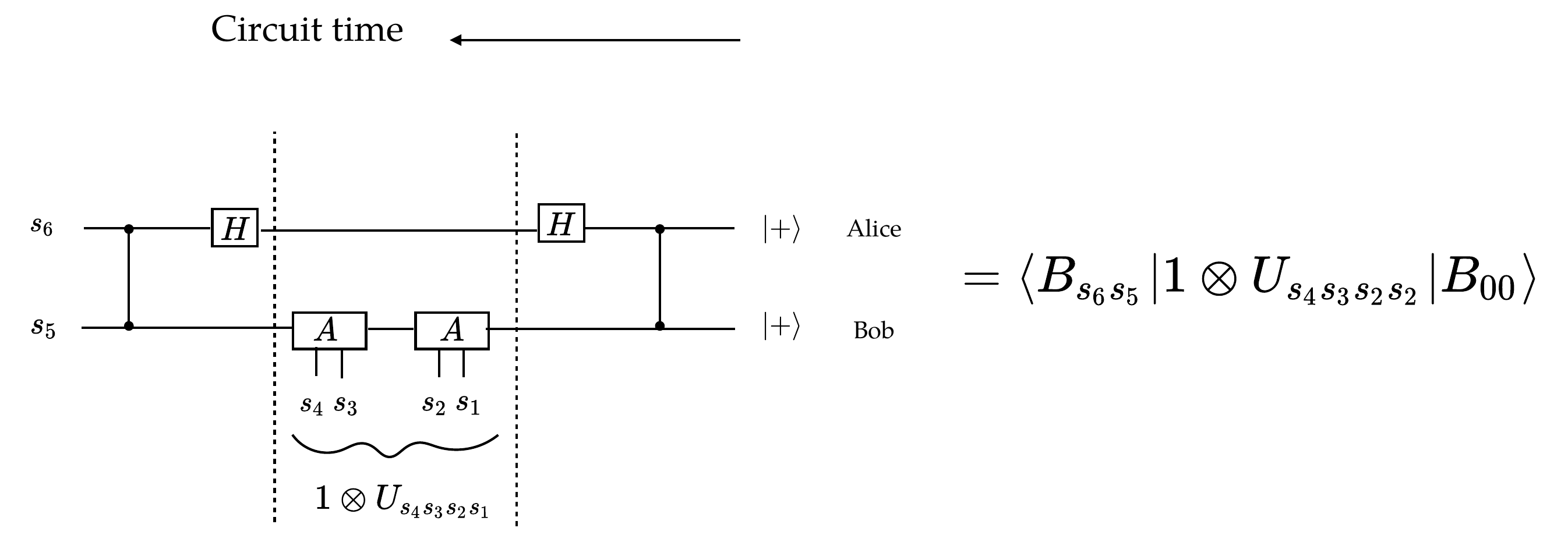}
\caption{The left diagram shows the MPS description of the cluster state on a periodic lattice of 6 qubits, corresponding to a state $\ket{1}$ with identify flux.   To see the corresponding circuit, we ``open" one of the MPS tensors to reveal a circuit on two qubits initialized on $\ket{+}$.  On the right, we interpret the same circuit as initializing a bell pair $\ket{B_{00}}$, which is subjected to  unitary $1\otimes U_{s_{4}s_{3}s_{2}s_{1}} $, followed by a measurement in the bell basis $\ket{B_{s_{1}s_{2}}} $}
\label{IDloop} 
\end{figure}

Our operational definition \eqref{hol} of entanglement holonomy involved a sequence of measurements, each of which induces a local teleportation of a logical qubit associated with edge modes living in $\mathcal{V}$.     However, the reader might wonder what happens to this logical  qubit when we make the final measurement in the sequence: how does this qubit ``annihilate" itself?   Here we explain the fate of this qubit by giving a circuit model description for teleportation in a closed loop.     This is illiustrated in figure \ref{IDloop}, where we consider the the simulation of the identity gate on a cluster state with 6 qubits. 
To obtain the circuit description, we open up one of the three MPS tensors using the circuit description in \ref{MPST}.    On the right of figure \eqref{IDloop}, we show how the measurement protocol described above effectively measures the amplitude $\braket{B_{s_{6}s_{5}} | 1\otimes U_{s_{4}s_{3}s_{2}s_{1}}| B_{00}}$  for the following process:   
\begin{enumerate}
\item a bell pair $\ket{B_{00}}$ is prepared by entangling two qubits A and B intialized in $\ket{+}$
\item The measurements of qubits 1-4 implements a  random unitary $U_{s_{4}s_{3}s_{2}s_{1}  }\sim  Z^{s_{1}+s_{3}} X^{s_{2}+s_{4}}$ on one member (say  B) of the bell pair
\item A bell measurement is made on the resulting entangled pair which projects onto $\ket{B_{s_{6}s_{5}}}$.   
\end{enumerate}

In the circuit model description, 
the final bell pair measurement is used to detect the unitary that was applied to qubit $B$, using qubit $A$ as a ``quantum" reference frame. 
In the circuit model description, the statement that the resource state $\ket{V_{g}=1}$ has a definitely holonomy translates into the fact that 
\begin{align}
    (s_{5},s_{6}) = ( s_{1}+s_{3},s_{2}+s_{4})
\end{align}
So that the last bell measurement determines the unitary implemented on the qubit:
\begin{align}
    U_{s_{4}s_{3}s_{2}s_{1} }  = Z^{s_{5}} X^{s_{6}} 
\end{align}
More generally, a nontrivial entanglement holonomy $V_{g}$ means that this relation is offset to become
\begin{align}
     U_{s_{4}s_{3}s_{2}s_{1} } V_{g} = Z^{s_{5}} X^{s_{6}} 
\end{align}
\section{Group labels for the temporally ordered measurement basis}
\paragraph{Measurement adaptations by symmetry}
Given a resource state  described by the a G-equivariant TQFT, which corresponds to an SPT phase protected by G, the measurement adaptations in MBQC can be implemented by the the adjoint representation of $G$, which acts on the ith site as
\begin{align} 
u(g_{i}) = l(g_{i} )r(g_{i})
\end{align} 
with $l(g_{i}), r(g_{i}) $ the representations of left and right multiplication described in the text.   The operation $u(g)$ can be used to  propagate the by products operators forward as show in the following figure: 

\begin{align}
\includegraphics[scale=.5]{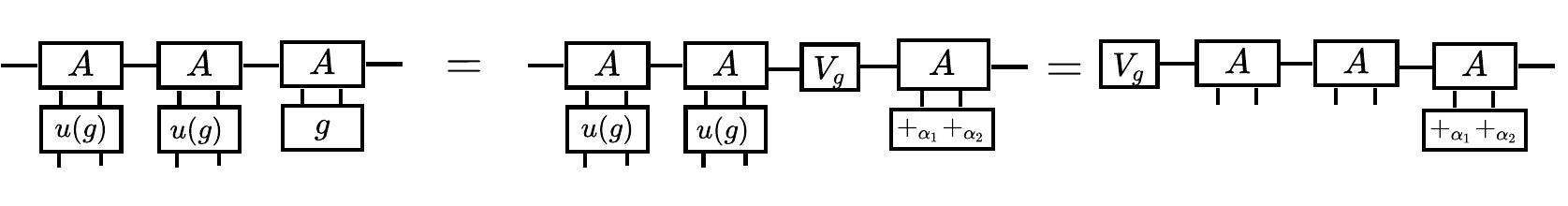}
\end{align} 

The total cumulative action of the measurement adaptation on  on the $j^{\rm th}$ block is:
\begin{align}
u(g_{\prec j}) &= 
X^{s_{2j-3} + \ldots + s_3 + s_1} \otimes
X^{s_{2j-2} + \ldots + s_4 + s_2}\nn
    g_{\prec j} &:= g_{1} g_{2} \ldots g_{j-1} 
\end{align}
This leads us to define an adapted reference basis given by :
\begin{equation}
 u(g_{\prec j}) 
\ket{+_{\alpha_{2j}}}
\otimes 
\ket{+_{\alpha_{2j-1}}}=\ket{+_{\big((-)^{q_{2j}-s_{2j-1}}\alpha_{2j}\big)}}
\otimes 
\ket{+_{\big((-)^{q_{2j-1}}\alpha_{2j-1}\big)}}
\label{reffi}
\end{equation}
Relative to this reference state, the measurement basis at the $j$th block 
is:
\begin{align}
\label{globalbasis} 
\ket{g_{j}}= l(g_{j}) u(g_{\prec j}) 
\ket{+_{\alpha_{2j}}}
\otimes 
\ket{+_{\alpha_{2j-1}}}
\end{align} 

In terms of the desired local unitary $U_{j}$  the adapted basis, the MPS tensors take the form 
\begin{align} \label{VA2}
    A_{g_{j}}=V_{g_{\prec i}} \left( V_{g_j} U_{j} \right)  V_{g_{\prec i}}^\dagger
\end{align}
\subsection{Stabilizers generate MBQC gauge transformations} \label{sec:stabgauge}
We check explicitly that the stabilizers \eqref{gauge} preserve the MBQC holonomies by applying them to the basis \eqref{globalbasis}:
\begin{align}
l_{i-1}(g) r_{i}(g)) \ket{h_{i}}_{U} \ket{h_{i-1}}_{U} &= r_{i}(g^{-1})  l(h_{i})  u(h_{\prec i}) \ket{+_{\alpha_{2i}}+_{\alpha_{2i-1}}} \otimes  l_{i-1}(g) \ket{h_{i-1}}_{U}\nn
&= l(h_{i})  r_{i}(g)  u(h_{\prec i}) \ket{+_{\alpha_{2i}}+_{\alpha_{2i-1}}} \otimes  \ket{g h_{i-1}}_{U}\nn
&= l(h_{i}g^{-1} )  r_{i}(g)  u(h_{\prec i} g^{-1}  ) \ket{+_{\alpha_{2i}}+_{\alpha_{2i-1}}} \otimes  \ket{ gh_{i-1}}_{U}\nn
&=l(h_{i}g^{-1} ) u(h_{\prec i}) \ket{+_{\alpha_{2i}}+_{\alpha_{2i-1}}} \otimes  \ket{ h'_{i-1}}_{U}\nn
&=  \ket{h'_{i}}_{U} \ket{h'_{i-1}}_{U}\end{align} ,
where $h'_{i} = h_{i} g^{-1} $ and $h'_{i-1} = g h_{i} $.  We see that the total holonomy of the two affected sites are preserved, i.e. $h'_{i}h'_{i-1} = h_{i} h_{i-1} $.

\bibliographystyle{utphys}
\bibliography{MBQC}

\end{document}